\documentclass[12pt,fleqn]{article}
\usepackage[utf8]{inputenc}
\usepackage{a4wide}

\usepackage{algpseudocode}
\usepackage{rotating}
\usepackage{caption}
\usepackage[ruled,vlined]{algorithm2e}
\usepackage{amsmath}
\usepackage[round]{natbib}
\usepackage{graphicx}   
\usepackage{enumitem}
\usepackage{mathtools}
\usepackage{bm}
\usepackage{appendix}
\usepackage{amsthm}
\usepackage{subcaption}
\newtheorem{theorem}{Theorem}[section]

\newtheorem{proposition}[theorem]{Proposition}
\usepackage{xcolor}
\usepackage{comment}
\usepackage[textsize=scriptsize]{todonotes}
\usepackage{ulem}
\usepackage{amsmath}
\usepackage{amssymb}
\usepackage{rotating}
\usepackage{pdfpages}
\usepackage[hidelinks]{hyperref}
\hypersetup{
    colorlinks,
    linkcolor={red!50!black},
    citecolor={blue!50!black},
    urlcolor={blue!80!black}
}

\usepackage{array}
\title{Fast approximate estimation of conditional Shapley values when using a linear explainer}
\author{Fredrik Lohne Aanes}
\date{July 2026}

\begin{document}
  
\maketitle

\abstract{In this paper, we develop three new methods, two approximate and one exact, for fast estimation of \textit{conditional} Shapley values when a linear regression model is used as the explainer. We apply constrained Gaussian Markov Random Field theory and sparse matrix algebra packages to estimate the coefficients of all submodels jointly and not sequentially. In the numerical case studies, our methods match the accuracy of the Sequential method in the shapr R-package while reducing the computation time from hours to minutes or seconds. Compared to the Iterative method in shapr, which samples coalitions until a convergence criterion is met, our methods estimate all $2^p$ coalitions when there are $p$ predictors present. When the iterative method requires all or a large fraction of the coalitions, our methods are substantially faster and with equal or better accuracy. When few coalitions suffice and the shapr iterative method is very fast, our methods deliver enumeration of all submodels, and in one case study over two million models are estimated within minutes. We recommend the exact method, since it has no tuning parameters and the three methods give comparable results.}

\section{Introduction}
We introduce three efficient procedures for the estimation of conditional Shapley values when we have mixed types of covariates. We use a linear explainer to obtain the Shapley values. The linear explainer fits multiple linear regression models with different subsets of covariates to the predictions of the model. This technique accounts for covariate dependence when estimating the Shapley values. However, when using exact calculations, the number of models that needs to be estimated is $2^p$ models, which becomes more and more computationally demanding  to estimate when the number of covariates, $p$, becomes large. In this paper we show that we can fit thousands of different linear regression models \textit{approximately} or exactly at the same time and do this relatively fast. We need to exploit the sparseness of the \textit{joint} precision matrix. Sparse matrix algebra exploits that the matrix involved has many 0s.

Our main contributions are
\begin{itemize}
    \item We suggest two approximate methods for estimating all submodels of a linear model fast. We provide convergence proofs, showing that the methods are convergent as the tuning parameter grows/shrinks.
    \item We suggest one fast and exact method for estimating all submodels of a linear model. 
    \item We provide several numerical studies comparing the output from our algorithms with the output from the shapr R-package available on CRAN. The estimated Shapley values by our new methods are equal to or better than the estimates from the shapr package, but our methods can be much faster.
\end{itemize}

In the first approximate method, one can fit submodels of a linear regression model approximately by adding large numbers to the diagonal of the precision (i.e. inverse covariance) matrix of the original model with all covariates. 

In the second method, we alter the design matrix by multiplying it with a projection matrix, so that the corrected model is distributed as an intrinsic Gaussian Markov Random Field. By proper treatment, we can approximately obtain the constrained parameter estimates. 

In the third and final method, we show that we can drop the 0-constrained parameters and only estimate the remaining parameters. The method is therefore without approximations.

For conditional Shapley values, the main package for obtaining Shapley value estimates is the shapr-package \citep{jullum2025shapr}. This package is therefore the natural competitor to our work. For the marginal Shapley values, there are numerous packages and software implementations available. As our goal is to estimate conditional Shapley values, we compare our work to shapr. 

There is a related method implemented in the leaps R package \citep{leapsPackage}. It performs best subset variable selection, which is a different problem. The underlying Fortran code (by Allan Miller) uses a branch-and-bound algorithm based on the original ideas by \cite{Furnival1974}. The package avoids estimating all submodels. In our problem, however, we need to estimate \textit{all} submodels and not only the best ones. 

The paper is structured as follows. In Section \ref{sec:BriefOverview} we give an introduction to Shapley values, in Section \ref{sec:C_GMRF} we introduce constrained Gaussian Markov Random Fields, which is the backbone of our methodology. In Section \ref{sec:NewMethod} we derive the results and introduce the new computational algorithms. In Section \ref{sec:numericalSection} we provide several numerical case studies. In Section \ref{sec:DiscussionConclusion} we provide a discussion and also a conclusion.

\section{Brief overview of Shapley values used to explain individual model predictions}\label{sec:BriefOverview}

The Shapley value \citep{Shapley1953} originates from game theory and is used to distribute payoff to the players of a game based on their contributions. In machine learning it is used to tell how much the different features contributed to a specific prediction of a model. We now formalize this. 

Assume we want to explain a prediction of a model $f(\bm d)$. The model is trained on
$\mathcal{D}=\{\bm d^{[i]},Y^{[i]}\}_{i=1}^{N_{train}}$, where $\bm d_{i}$ is a $p$-dimensional feature vector, $Y^{[i]}$ is a univariate response and the number of training observations is $N_{train}$. The prediction $\hat{Y}=f(\bm d)$ for a feature vector $\bm d=\bm d^*$ is to be explained by using Shapley values. Denote the Shapley value corresponding to feature $i$ by $\phi_i$. The Shapley value vector $\bm \phi$ is a break-down of the prediction into the contribution of each of the predictors. This means $\phi_i$ tells us how much the $i$-th feature contributed to the specific prediction of a model in a positive (positive value) or negative way (negative value).

To calculate $\bm\phi$ we need to define what is called the contribution function, which should resemble the value of $f(\bm d^*
)$ when only the features in coalition $S$ are known.
We use the contribution function \citep{Lundberg2017}
\begin{equation}
\label{eq:shapleyConditional}
    v(S)=E(f(\bm d)|\bm d_s=\bm d_s^*),
\end{equation}
which means we consider the expected value of $f(\bm d)$ conditioned on the features in $S$ taking on the values $\bm d
_s^*$. For instance when $S$ consists of the first two features, then the conditional expectation is conditional on these two features only and at the specified values. We must calculate the contributions for each possible configuration of covariates to be able to find the Shapley values. For $p$ features, there are $2^p$ possible subsets, which means the procedure becomes intractable when there are many features. 

We use the so-called KernelSHAP procedure so that we estimate $\bm \phi$ by solving a linear weighted least squares problem. The Shapley values $\bm\phi$ are obtained by solving the system of equations 
\begin{equation}
\label{eq:kernelShap}
    \bm B^T\bm W\bm B\bm\phi  = \bm  B^T \bm W \bm v,
\end{equation}
here $\bm v$ is the vector of expected contribution function of length $2^p$, $\bm B$ is a $2^p\times (p+1)$ binary matrix (0 and 1), where the first column is a column vector of 1s and the $i$-th element in the j-th row is 1 if covariate $i$ is  in the set $S$ corresponding to the $j$-th value of contribution function and otherwise 0, and the weight matrix $\bm W$ is a diagonal matrix  $2^p\times 2^p$ with elements $k(p,|S|)=(p-1)/\binom{p}{|S|}|S|(p-|S|)
$. Here $|S|$ is the number of covariates in the set $S$. We set $k(p,0)$ and $k(p,p)$ to $10^6$, which is also done in the shapr-package \citep{jullum2025shapr}.

When the number of covariates becomes large, one can also sample the subsets (with replacement), which are also called coalitions, used to estimate $\bm \phi$. One uses the Shapley weights $k(p,|S|)$ as sampling weights and always includes the empty and full coalitions in the estimation.  

There are mainly two ways to estimate $v(S)$, Monte Carlo methods and regression methods. The Monte Carlo methods use sampling to estimate the conditional expectation. The regression paradigm directly fits a regression model that predicts $v(S)$. We focus only on the regression paradigm in our work.

There is also a popular Shapley value called the marginal Shapley value. The term \textit{marginal} stems from the fact that it does not take into account feature dependence in the estimation phase. The first paper to take into account feature dependence is \cite{Aas2021}, and the corresponding Shapley value is the conditional one. The shap package \citep{Lundberg2017} is a highly popular package used to estimate (mainly) marginal Shapley values, while shapr \citep{jullum2025shapr} estimates conditional Shapley values. The shapr-package can be slower since it takes into account feature dependence.

More details on Shapley value estimation and background theory for prediction models is given in e.g. \cite{Lundberg2017}, \cite{Aas2021} or \cite{olsen2023comparative}.

We compare our new methods with two methods in the shapr-package. The first method, which we call the Sequential estimation method, is called the \textit{Regression separate method} in their work. This method estimates Shapley values using all coalitions. However, instead of estimating each coalition in a for-loop, the package uses parallelization. One can force the shapr package to perform calculation in serial mode.

The authors of the shapr package have also developed an iterative KernelSHAP method, where coalitions are added sequentially to the estimation method until a convergence criterion has been fulfilled. We combine this procedure with the Regression separate method from the same package, using already built-in procedures. The iterative procedure uses bootstrapping \citep{Efron1979}  to estimate the variance of the Shapley value estimates, and when the bootstrap variance becomes small, i.e., based on the criterion mentioned above, the package stops.

\section{Constrained Gaussian Markov Random Fields}\label{sec:C_GMRF}
In this section we provide a short summary of constrained Gaussian Markov Random Fields. The theory is needed in the development of the new theory.  

A multivariate random variable $\bm z$ of length $n_z$ is said to be a Gaussian Markov Random Field if the density is given by \citep[p.~22]{rue2005gaussian}
\begin{equation}
    f(\bm z)=(2\pi)^{-n_z/2}|\bm Q|^{1/2}\exp\big(-\frac{1}{2}(\bm z-\bm \mu)^T\bm Q (\bm z-\bm \mu )\big),
\end{equation}
where $\bm \mu$ is the mean vector and $\bm Q$ is the precision matrix. In several cases $\bm Q=\bm R\tau$, where $\bm R$ is the so-called structure matrix and both the structure and values are fixed. In linear regression $\bm R=\bm X^T\bm X$, where $\bm X$ is the design matrix. The parameter $\tau$ is a hyper-parameter called the precision parameter, and is connected to the variance parameter by $\tau =1/\sigma^2$. 
It is well-known that the parameter estimators in the multivariate linear regression are Gaussian if the observations are assumed to be Gaussian. The estimated mean vector is given by the solution of the Normal equations, i.e.
\begin{equation}
\label{eq:NormaEq}
    \bm \mu =(\bm X^T \bm X)^{-1}\bm X^T \bm Y,
\end{equation}
and the precision matrix is given by $\bm Q=\bm X^T\bm  X\tau$.

The distribution of $\bm z$ under linear constraints $\bm A \bm z =\bm 0$ can be obtained by considering the distribution of $(\bm z,\bm A\bm z)$ and then conditioning on $\bm A\bm z =\bm 0$ using standard multivariate normal distribution theory. One obtains that
\begin{equation}
\label{eq:sigmaKrig}
    \bm \Sigma^*_A = \bm Q^{-1}-\bm Q^{-1} \bm A^T(\bm A\bm Q^{-1}\bm A^T)^{-1}\bm A\bm Q^{-1},
\end{equation}
and 
\begin{equation}
    \bm\mu^*_A=\bm\mu -\bm Q^{-1} \bm A^T(\bm A\bm Q^{-1}\bm A^T)^{-1}\bm A\bm \mu.   
\end{equation}
By comparing the two above equations, we have that 
\begin{equation}
\label{eq:MeanCorrectionEquation}
    \bm \mu^*_A=\bm\Sigma^*_A \bm Q\bm\mu.
\end{equation}
We call the above equations the correction equations, as we correct the original parameters (mean vector and covariance matrix) for the constraints.

In the regression setting, we impose several constraints. Denote the i-th constraint by $\bm A_i$. This is a row vector where the i-th element is 1 if we set the regression coefficient to 0 in the model and 0 elsewhere on that row. We collect all such vectors in the $\bm A$ constraint matrix. This is equivalent to imposing the so-called corner constraint. Here we split the $\bm z$ vector in two, $\bm z_B$ and $\bm z_C$ and condition the vector $\bm z_B$ on $\bm z_C=0$. Partition the precision matrix as follows
\begin{equation}
\label{eq:QPartition}
\begin{bmatrix}
   \bm  Q_B & \bm  Q_{BC}\\
    \bm Q_{CB} & \bm Q_{C}
\end{bmatrix}.
\end{equation} From Theorem 2.5 in \cite{rue2005gaussian}, the precision matrix under the corner constraint is $\bm Q_B$. This means we can obtain the precision matrix of a smaller model by removing columns from the original model matrix $\bm X$. 

\section{Methodology and computational algorithms}\label{sec:NewMethod}
We now derive three methods, two approximate and one exact method. We propose the Diagonal correction method in Section \ref{sec:DiagonalMethod}, the Projection method in Section \ref{sec:ProjectionMethod} and the Exact transformation method in Section \ref{sec:ExactMethod}. In Section \ref{sec:Algorithms} we propose algorithms using the newly developed methods that are used in the numerical section.

\subsection{The Diagonal correction method}
\label{sec:DiagonalMethod}
We now develop a method where we obtain a submodel by adding large numbers to the diagonal of the precision matrix. 

The next result is due to the Woodbury matrix inversion formula.
\begin{proposition}
\label{pr:CondVar}
    The conditional covariance matrix, $\bm\Sigma^*$, when imposing constraints $\bm A\bm z=\bm 0$ in the linear regression estimation problem,  can be approximated with
    \begin{equation}
      \bm\Sigma^D_A=(\bm Q+\bm A^T\bm A\bm\kappa)^{-1}=\bm\Sigma-\bm\Sigma\bm A^T (\bm A\bm\Sigma \bm A^T+\bm I/\kappa)^{-1}\bm A\bm\Sigma \stackrel{\kappa\rightarrow \infty}{\rightarrow}  \bm \Sigma_A^*.
    \end{equation}
\end{proposition}
\begin{proof}
 By applying the Woodbury identity, see \cite{woodbury1950inverting}, using $\bm Q, \bm A^T,\bm A$ and $\bm I\kappa$ as input matrices, we obtain  
\begin{equation*}
\label{eq:main}
\bm\Sigma_A^D=(\bm Q+\bm A^T\bm A\kappa)^{-1}=\bm\Sigma-\bm\Sigma\bm A^T (\bm A\bm\Sigma \bm A^T+\bm I/\kappa)^{-1}\bm A\bm\Sigma \stackrel{\kappa\rightarrow\infty}{\rightarrow}\bm\Sigma^*_A,
\end{equation*}
where $\bm\Sigma^*_A$ is given in Equation \eqref{eq:sigmaKrig}.
\end{proof}
We note that $\bm A\bm\Sigma\bm A^T$ has been replaced by $\bm A\bm\Sigma \bm A^T+\bm I/\kappa$ in the correction equation for $\bm\Sigma^*$, which means that $\bm\Sigma_A^D$ is an approximation of $\bm\Sigma^*$ and improves, in theory, when $\kappa$ increases. However, if $\kappa$ is chosen too large, it will become numerically infinite due to the finite machine-precision. This means we can approximate the corrected covariance matrix by using $\bm Q+\bm A^T\bm A\kappa$ as precision matrix, where $\bm A$ is the constraint matrix of a single model. Our goal is to estimate the coefficients of all submodels of a linear model. We make a joint constraint matrix $\mathcal{\bm A}$, where we pad individual constraint matrices with $0$'s. For instance if we consider a simpler setup with two covariates, then there are four models, and $\mathcal{\bm A}$ could take the following form:
\begin{align*}
    \mathcal{\bm A}=\begin{bmatrix}
0 & 1 & 0 & 0  & 0 & 0 & 0 & 0 & 0 & 0 & 0 & 0\\
0 & 0 & 1 & 0  & 0 & 0 & 0 & 0 & 0 & 0 & 0 & 0\\
0 & 0 & 0 & 0  & 1 & 0 & 0 & 0 & 0 & 0 & 0 & 0\\
0 & 0 & 0 & 0  & 0 & 0 & 0 & 0 & 1 & 0 & 0 & 0\\
    \end{bmatrix}
\end{align*}
The constraint matrix has dimensions $n_\mathcal{A}\times (|\bm z|\cdot 2^p)$. Here $n_\mathcal{A}$ is the number of rows in the constraint matrix, which is 4 in the example. Furthermore, $|\bm z|$ is the length of the linear predictor, which is 3 in our small example. And, finally, $p$ is the number of predictors. Therefore, the constraint matrix has dimensions $4\times 12$ in our example. The first two rows from the top specify a model where all coefficients are set to zero, except for the intercept. The following row specifies a model where the first coefficient is set to zero only. The next row specifies a model where the last parameter coefficient is set to zero. There are no constraints on the fourth and final model. Each row in $\mathcal{A}$ gives a constraint on one model only. 

In the next proposition we derive the Diagonal correction method and prove that the error converges to 0. 
\begin{proposition}
\label{pr:errorbound}
 \begin{enumerate}[label=\textup{(\alph*)}, leftmargin=*]
\item Denote the precision matrix of the full model by $\bm Q=\bm X^T\bm X\tau$ and the constraint matrix by $\bm A$. We find the approximate solution by calculating 
\begin{equation}
   \bm \mu_A^D= \bm\Sigma^D_A\bm Q\bm \mu\rightarrow\bm\Sigma_A^*\bm Q\bm\mu=\bm \mu^*_A, 
\end{equation} where $\bm\Sigma^D_A$ is defined in Proposition \ref{pr:CondVar} and $\bm\mu_A^D$ are the regression coefficients using the Diagonal correction method. The vector $\bm \mu_A^*$ is the regression parameter estimates found by solving Equation \ref{eq:MeanCorrectionEquation}.
\item The error between the true solution, $\bm \mu_A^*$ and the approximated solution, $\bm\mu_A^D$, is for all values of $\kappa>0$ 
\begin{align*}
\bm\mu^D_A-\bm\mu^*_A&=\bm\Sigma\bm A^T((\bm A\bm\Sigma\bm A^T)^{-1}-(\bm A\bm\Sigma\bm A^T+\bm I/\kappa)^{-1})\bm A\bm\mu\\&=\bm \Sigma\bm A^T((\bm A\bm\Sigma\bm A^T)^{-1}((\bm A\bm\Sigma\bm A^T)^{-1}+\bm I\kappa)^{-1}(\bm A\bm\Sigma\bm A^T)^{-1})\bm A\bm \mu\\& \stackrel{\kappa \rightarrow\infty} {\rightarrow}\bm 0. 
\end{align*}
\item If we let $\bm Q=\bm X^T\bm X$, let $\lambda_{max}$ denote the largest eigenvalue of the precision matrix of the original model (i.e. $\bm Q$), and let $\eta_k$ denote any eigenvalue of the precision matrix of a submodel, then \begin{align*}
\lambda_{max}\geq \eta_k,
\end{align*}
for all $k$. 
The fixed tuning parameter $\kappa$ should be chosen sufficiently large compared to $\lambda_{max}$. 
\item Assume we want to estimate the coefficients for all submodels of a linear model. Collect the constraints in the matrix $\mathcal{\bm A}$. One pads the constraint matrix for each submodel with 0's. Assume $\bm Q_F=\bm I_{2^p}\otimes \bm X^T\bm X$, where $\bm I_{2^p}$ is the $2^p\times 2^p$ identity matrix. Let also $\bm m$ be a vector where $\bm X^T\bm f$ is repeated $2^p$ times. Then the approximate parameter estimates given by the Diagonal method are given by
\begin{align}
\label{eq:DcorrectionEquation}
    \bm\mu^D_\mathcal{A} = (\bm Q_F+ \kappa \mathcal{\bm A}^T\mathcal{\bm A})^{-1}\bm m \stackrel{\kappa\rightarrow \infty}{\rightarrow}\bm\mu_{\mathcal{A}}^*,
\end{align}
\end{enumerate}
where $\bm\mu_{\mathcal{A}}^*$ is the vector with regression parameter estimates obtained by solving the Normal equations for each submodel, so that the approximation error converges to $\bm 0$. 
\end{proposition}
\begin{proof}
\begin{enumerate}[label=(\alph*)]
\item The approximate result follows by applying Equation \ref{eq:MeanCorrectionEquation} and by replacing $\bm\Sigma_A^*$ with $\bm\Sigma^D_A$ from Proposition \ref{pr:CondVar}. Due to cancellation of $\tau$, we can pick  $\tau =1$ when calculating $\bm \mu^D_A$.
\item 
By applying Proposition \ref{pr:CondVar}, we get \begin{align*}
\bm \mu_A^D-\bm\mu_A^* &=\bm \Sigma^D_A\bm Q\bm\mu-\bm \Sigma^*\bm Q\bm\mu\\&=-\ \bm\Sigma\bm A^T(\bm A\bm\Sigma\bm A^T+\bm I/\kappa)^{-1}\bm A\bm\mu+\ \bm\Sigma\bm A^T(\bm A\bm\Sigma\bm A^T)^{-1}\bm A\bm\mu\\&=\bm\Sigma\bm A^T((\bm A\bm\Sigma\bm A^T)^{-1}-(\bm A\bm\Sigma\bm A^T+\bm I/\kappa)^{-1})\bm A\bm\mu
\\&=\bm \Sigma\bm A^T((\bm A\bm\Sigma\bm A^T)^{-1}-(\bm A\bm\Sigma\bm A^T)^{-1}+\\&(\bm A\bm\Sigma\bm A^T)^{-1}((\bm A\bm\Sigma\bm A^T)^{-1}+\bm I\kappa)^{-1}(\bm A\bm\Sigma\bm A^T)^{-1})\bm A\bm \mu \\& =\bm \Sigma\bm A^T((\bm A\bm\Sigma\bm A^T)^{-1}((\bm A\bm\Sigma\bm A^T)^{-1}+\bm I\kappa)^{-1}(\bm A\bm\Sigma\bm A^T)^{-1})\bm A\bm \mu,
\end{align*}
where we have applied the Woodbury identity on $(\bm A\bm\Sigma \bm A^T+\bm I/\kappa)^{-1}$.
\item The covariance matrix of the full model is given by $\bm\Sigma =(\tau \bm Q)^{-1}$. We can pick $\tau =1$ due to cancellation. Due to the Poincaré separation theorem, also known as the Cauchy interlacing theorem, the smallest eigenvalue of $\bm \Sigma$ is equal to or smaller than any of the eigenvalues of the covariance matrix of any submodel $\bm A\bm \Sigma\bm A^T$, where the $\bm A$ matrix has orthogonal rows. Since the largest eigenvalue of $\bm Q$ is equal to the reciprocal of the smallest eigenvalue of $\bm \Sigma$, then $\lambda_{max}\geq \eta_k$. We have fixed $\tau =1$ when considering the covariance matrix of any submodel also. The reason is that we consider calculating the mean parameter, where no $\tau$ enters the calculations. From part (b) of this proposition,
\begin{align*}\bm \mu_A^D-\bm\mu_A^*  = \bm \Sigma\bm A^T((\bm A\bm\Sigma\bm A^T)^{-1}((\bm A\bm\Sigma\bm A^T)^{-1}+\bm I\kappa)^{-1}(\bm A\bm\Sigma\bm A^T)^{-1})\bm A\bm \mu,
\end{align*}
so that if $\kappa$ in $\kappa\bm I$ is chosen to be much larger than the largest eigenvalue of $(\bm A\bm\Sigma\bm A^T)^{-1}$, we obtain approximate convergence. The matrix $(\bm A\bm\Sigma\bm A^T)^{-1}$ is the precision matrix of the reduced model. We should therefore pick $\kappa$ sufficiently large compared to $\lambda_{max}$.
\item The result follows due to the block-diagonal structure in $\bm Q_F$, that $\mathcal{\bm A}$ has orthogonal row blocks and that the Diagonal method is valid for each submodel. 
\end{enumerate}
\end{proof}
From Proposition \ref{pr:errorbound} we have a method for calculating a candidate solution for the regression coefficients in a constrained linear regression. And we can bound the error by choosing a matrix norm, e.g. the Frobenius norm or another matrix norm. We also note that $\bm Q+\bm A^T\bm A\kappa$ is always mathematically positive-definite, assuming that $\bm Q$ is positive definite ($\bm X$ has full rank).  The reason is that $\bm e^T(\bm A^T\bm A\kappa)\bm e\geq 0$ for all vectors $\bm e$ since  $\kappa$ is set to a large positive fixed value. The tuning parameter $\kappa$ needs to be chosen large enough. We suggest to take the eigendecomposition of the original model with all covariates, i.e. $\bm Q$ and find the largest eigenvalue $\lambda_{max}$. Then we can pick $\kappa = \lambda_{max}\cdot 10^{5}$. 

In the Diagonal correction method we force approximately 0-constrained parameters to be 0 by adding a large value to each diagonal element of precision matrix corresponding to a 0-constrained parameter. The Woodbury identity makes this a valid procedure. When we examine the output of an implementation of the method, 0-constrained parameters will be small in value.   
\subsection{The Projection method}
\label{sec:ProjectionMethod}
Next, we present another method for fast calculation of the mean parameters in linear regression. We name it the Projection method.
\begin{proposition}
\label{pr:Method2}
\begin{enumerate}[label=(\alph*)]

\item Assume we have $u$ orthonormal constraints and that we collect them in the constraint matrix $\bm A$. Define $\bm Z=\bm I-\bm A^T\bm A$ and denote by $\bm Q$ the precision matrix of the full model, then $\bm Z\bm Q\bm Z$ is the precision matrix of the constrained model. 
 \item By adding $\bm I \varepsilon$ to $\bm Z\bm Q\bm Z$, inverting and applying $\bm Z$, we obtain
\begin{equation*}
    \bm Z(\bm Z\bm Q\bm Z+\varepsilon \bm I)^{-1}\bm Z\stackrel{\varepsilon\rightarrow 0^+}{\rightarrow} \bm \Sigma^*_A,
\end{equation*}
where $\bm\Sigma^*_A$ is given in Equation \eqref{eq:sigmaKrig}.
\item The approximation error converges to $\bm 0$ as $\varepsilon\rightarrow 0^+$, i.e. 
\begin{align*}
    \bm\mu_A^P-\bm\mu_A^* \stackrel{\varepsilon\rightarrow 0^+}{\rightarrow} \bm 0,
\end{align*}
where $\bm \mu_A^P=\bm Z(\bm Z\bm Q\bm Z+\varepsilon \bm I)^{-1}\bm Z\bm Q\bm \mu$ is the estimates of the regression coefficients under the Projection method while $\bm \mu_A^*$ denotes the regression parameter estimates by solving Equation \ref{eq:MeanCorrectionEquation}.
\item Assume we want to estimate the coefficients for all submodels of a linear model. Collect the constraints in the matrix $\mathcal{\bm A}$. One pads the constraint matrices for each submodel with 0's. Assume $\bm Q_F=\bm I_{2^p}\otimes \bm X^T\bm X$, where $\bm I_{2^p}$ is the $2^p\times 2^p$ identity matrix. Let also $\bm m$ be a vector where $\bm X^T\bm f$ is repeated $2^p$ times. Calculate $\bm Z =\bm I-\mathcal{\bm A}^T\mathcal{\bm A}$, we then have:
\begin{align}
\label{eq:PcorrectionEquation}
    \bm\mu^P_{\mathcal{A}}=\bm Z(\bm Z\bm Q_F\bm Z+\varepsilon \bm I)^{-1}\bm Z\bm m\stackrel{\varepsilon\rightarrow 0^+}{\rightarrow}\bm\mu_\mathcal{A}^*,
\end{align}
where $\bm\mu^P_{\mathcal{A}}$ denotes the estimates under the Projection method, while $\bm\mu_\mathcal{A}^*$ is the vector with regression parameter estimates obtained by solving the Normal equations for each submodel. 
 \end{enumerate}
\end{proposition}
\begin{proof}
\begin{enumerate}[label=(\alph*)]
\item To establish that $\bm Z\bm Q\bm Z$ is the pseudo-inverse of $\bm \Sigma^*$ we have to verify the Moore-Penrose conditions see e.g. \cite{Penrose_1955}: (1) $\bm M\bm M^*\bm M=\bm M$, where $\bm M^*$ is the Moore-Penrose pseudo-inverse of $\bm M$, (2) $\bm M^*\bm M\bm M^*=\bm M^*$, (3) $(\bm M^*\bm M)^T=\bm M^*\bm M$ and (4)  $(\bm M\bm M^*)^T=\bm M\bm M^*$. 

In the first condition, we have to show that $\bm Z\bm Q\bm Z\bm \Sigma_A^*\bm Z\bm Q\bm Z=\bm Z\bm Q\bm Z$. We decompose the matrix product into two terms:
\begin{align*}
\bm Z\bm Q\bm Z\bm \Sigma_A^*\bm Z\bm Q\bm Z & =\bm Z\bm Q\bm Z\bm \Sigma\bm Z\bm Q\bm Z -\bm Z\bm Q\bm Z \bm\Sigma \bm A^T(\bm A\bm\Sigma\bm A^T)^{-1}\bm A\bm\Sigma\bm Z\bm Q\bm Z\\ & =(1)-(2).
\end{align*}

In the first term, we use that $\bm Z= \bm I-\bm A^T\bm A$ and that  $\bm Z$ is a projection matrix (which means that $\bm Z^2=\bm Z$). We obtain
\begin{align*}
(1) &= \bm Z\bm Q\bm Z\bm\Sigma \bm Z\bm Q\bm Z=\bm Z\bm Q\bm [\bm I -\bm A^T\bm A]\bm \Sigma\bm Z\bm  Q\bm Z\\ & = \bm Z\bm Z\bm Q\bm Z-\bm Z\bm Q\bm A^T\bm A\bm\Sigma\bm Z\bm Q\bm Z \\&= \bm Z\bm Q\bm Z-\bm Z\bm Q\bm A^T\bm A\bm\Sigma\bm Z\bm Q\bm Z.
\end{align*}

In the second term, we insert the expression for $\bm Z=\bm I-\bm A^T\bm A.$
\begin{align*}
    (2) &= \bm Z\bm Q[\bm I-\bm A^T\bm A]\bm\Sigma \bm A^T(\bm A\bm\Sigma\bm A^T)^{-1}\bm A\bm\Sigma\bm Z\bm Q\bm Z \\ &=(\bm Z\bm A^T-\bm Z\bm Q\bm A^T\bm A\bm\Sigma\bm A^T)(\bm A\bm\Sigma\bm A^T)^{-1}\bm A\bm \Sigma\bm Z\bm Q\bm Z
    \\&=\bm 0-\bm Z\bm Q\bm A^T\bm A\bm \Sigma\bm Z\bm Q\bm Z,
\end{align*}
where we have used that $\bm Z\bm A^T=\bm 0$.

In total, we obtain
\begin{align*}
    (1)-(2) &=\bm Z \bm Q \bm Z-\bm Z\bm Q\bm A^T\bm A\bm \Sigma\bm Z\bm Q\bm Z -[-\bm Z\bm Q\bm A^T\bm A\bm \Sigma\bm Z\bm Q\bm Z] \\ & =\bm Z\bm Q\bm Z,
\end{align*}
which is what we wanted to show for condition (1). 

Next, we prove the second condition. We get
\begin{align*}
\bm\Sigma^*_A\bm Z\bm Q\bm Z\bm \Sigma^*_A =&\bm\Sigma^*_A[\bm I-\bm A^T\bm A]\bm Q[\bm I-\bm A^T\bm A]\bm\Sigma^*_A\ =\bm\Sigma^*_A \bm Q \bm\Sigma^*_A \\=&(\bm I-\bm \Sigma \bm A ^T(\bm A\bm \Sigma\bm A^T)^{-1}\bm A)\bm \Sigma_A^*=\bm\Sigma_A^*, 
\end{align*}
as $\bm A\bm\Sigma^*_A=\bm 0$ and $\bm A^T\bm A\bm \Sigma_A^*=\bm \Sigma_A^*\bm A^T\bm A\bm=\bm 0$. We have now proved the second condition.

Next we consider the third condition. We first consider $\bm M^*\bm M$. We obtain
\begin{align*}
    \bm\Sigma_A^*\bm Z\bm Q\bm Z &=
\bm\Sigma_A^*[\bm I-\bm A^T\bm A] \bm Q\bm Z =\bm\Sigma_A^*\bm Q\bm Z =(\bm I-\bm\Sigma\bm A^T(\bm A\bm \Sigma\bm A^T)^{-1}\bm A)\bm Z \\ &=\bm I-\bm A^T\bm A-\bm\Sigma\bm A^T(\bm A\bm \Sigma\bm A^T)^{-1}\bm A+\bm\Sigma\bm A^T(\bm A\bm \Sigma\bm A^T)^{-1}\bm A\bm A^T\bm A\\ &=\bm I-\bm A^T\bm A,
\end{align*}
as $\bm A\bm A^T$ is the identity matrix and $\bm \Sigma^*_A\bm A^T=\bm 0$. Due to symmetry, the third condition is satisfied. 

We finally prove that the fourth condition is satisfied. We first consider $\bm M\bm M^*$. We obtain
\begin{align*}
    \bm Z\bm Q\bm Z \bm\Sigma_A^*&=\bm Z\bm Q[\bm I-\bm A^T\bm A]  \bm\Sigma_A^*=\bm Z\bm Q\bm \Sigma_A^* =\bm Z(\bm I-\bm A^T(\bm A\bm\Sigma\bm A^T)^{-1}\bm A\bm\Sigma)\\ &=\bm I-\bm A^T\bm A-\bm A^T(\bm A\bm\Sigma\bm A^T)^{-1}\bm A\bm \Sigma+\bm A^T\bm A\bm A^T(\bm A\bm\Sigma\bm A^T)^{-1}\bm A\bm \Sigma \\ &=\bm I-\bm A^T\bm A,
\end{align*}
since $\bm A\bm A^T$ is the identity matrix and $\bm A\bm \Sigma^*_A=\bm 0$. Due to symmetry, the fourth condition is satisfied. We have now proved that all four conditions are satisfied. Our proof is therefore complete. 
\item From part (a) of this proposition we know that $\bm Z\bm Q\bm Z$ is the correct precision matrix of the conditional model. We have that the rank of $\bm Z\bm Q\bm Z$ is $n-u$. The eigendecomposition is $\bm Z\bm Q\bm Z=\sum_{i=1}^{n-u}\lambda_i \bm c_i\bm c_i^T$. Since we add $\varepsilon\bm I$ to this matrix, we obtain the additional term $\bm A^T\bm A\varepsilon$. We get
\begin{align*}
\bm Z(\bm Z\bm Q\bm Z+\varepsilon \bm I)^{-1}\bm Z&=\bm Z (\sum_{i=1}^{n-u}\frac{1}{\lambda_i+\varepsilon}\bm c_i\bm c_i^T +\frac{1}{\varepsilon}\bm A^T\bm A)\bm Z \\ &=\bm Z (\sum_{i=1}^{n-u}\frac{1}{\lambda_i+\varepsilon}\bm c_i\bm c_i^T)\bm Z \stackrel{\varepsilon\rightarrow 0^+}{\rightarrow}\bm\Sigma^*_A,
\end{align*}
where $\bm \Sigma^*_A$ is defined in Equation \ref{eq:sigmaKrig}. The $\bm Z$-matrix removes the contribution in the $\bm A$-directions.   
\item We insert the expression for $\bm\mu_A^P$ from part (c) of this proposition and also insert the expression for $\bm\mu_A^*$ from Equation \eqref{eq:MeanCorrectionEquation} into the error expression and obtain:
\begin{align*}
    \bm\mu_A^P-\bm\mu_A^* =  \bm Z(\bm Z\bm Q\bm Z+\varepsilon \bm I)^{-1}\bm Z\bm Q\bm\mu-\bm \Sigma_A^*\bm Q\bm \mu \stackrel{\varepsilon\rightarrow 0^+}{\rightarrow} \bm \Sigma_A^*\bm Q\bm\mu-\bm\Sigma_A^*\bm Q\bm\mu =\bm 0.
\end{align*}
\item The result follows since the Projection method is correct for each submodel and we have independence between different models due to the block diagonal structure in $\bm Z\bm Q_F\bm Z$. As we have convergence for each submodel, the error also approaches 0 as $\varepsilon\rightarrow 0^+$ when considering $\mathcal{\bm A}$.   
\end{enumerate}
\end{proof}
In the Projection method we project away the 0-constrained parameters by applying $\bm Z=\bm I -\mathcal{A}^T\mathcal{A}$ to the global precision matrix. We then obtain the correct conditional precision matrix. Technically, the model with $\bm Z\bm Q\bm Z$ as precision matrix, is an intrinsic Gaussian Markov Random Field. The reason is that the precision matrix is not of full rank. See Chapter 3 in \cite{rue2005gaussian} for more details. We make the precision matrix positive definite by adding $\bm I\varepsilon$ for small $\varepsilon$. It could be chosen to be e.g. $10^{-5}$. If the smallest eigenvalue of $\bm X^T\bm X$ is smaller than $10^{-15}$, it indicates that the matrix is close to numerical singular. Care must then be taken. The rule is therefore to pick $\varepsilon$ small relative to the smallest eigenvalue of $\bm Q$, but large enough for stable Cholesky decomposition. We have suggested to use $\varepsilon =10^{-5}$, but one could also use $\varepsilon =10^{-10}$. By adding $\varepsilon \bm I$ to the precision matrix, we add $\varepsilon$ to each eigenvalue. It follows that we do not alter the other eigenvalues that much if we make the suggested choice for $\varepsilon$.  

\subsection{The Exact transformation method}
\label{sec:ExactMethod}
In the previously introduced Diagonal correction method we force parameters that should be 0 to be approximately 0. In the Projection method we introduce the $\varepsilon$ parameter, which introduces an approximation error. We now propose an exact method with no approximation error. In this method, instead of working in the full parameter dimension, we transform to a lower dimensional space where only non-zero elements are estimated. Coefficients which are known to be 0, are dropped from the model estimation. Let $P_F$ denote the size of the full parameter space and let $P_R$ denote the size of the parameter space with only non-zero parameters. It follows that $P_R<P_F$, and that $P_F=2^p\cdot n_z $, where $p$ is the number of covariates and $n_z$ is the length of the parameter vector of the original unconstrained model including the intercept. We introduce a mapping matrix $\bm E$. The matrix $\bm E$ has row size $P_R$ and column size $P_F$. Element $E_{i,j}$ equals 1 if element $j$ of the parameter vector is to be estimated. Which model row $i$ corresponds to is determined by the binary $\bm B$-matrix defined in Section \ref{sec:BriefOverview}. For example, consider the two first models defined in $\bm B$ (which corresponds to the first two rows in $\bm B$). Assume there are 5 covariates, not including the intercept. Also, assume that in the first model only the intercept is estimated, while the remaining coefficients are set to zero, then $E_{1,1}=1$. Assume that in the second model the intercept and the first continuous covariate are estimated, then $E_{2,7}=1$ and $E_{3,8}=1$. If the first covariate had been categorical, we would need to use several rows. The remaining coefficients and elements in $\bm E$ follow the same pattern. In the next proposition we present the method, which we name the Exact transformation method.        
\begin{proposition}
In the estimation of regression coefficients for all submodels of a linear model using the Exact transformation method, let $\bm E$ denote the mapping matrix. Denote by $\bm Q_F=\bm I_{2^p}\otimes \bm Q=\bm I_{2^p}\otimes \bm X^T\bm X$, where $\bm I_{2^p}$ denotes the $2^p\times 2^p$ identity matrix and let $\bm m$ be $\bm X^T\bm f$ repeated $2^p$ times, stacked on top of each other. The estimates of the regression parameters that are not constrained to be zero, are given by
\begin{align}
\label{eq:EcorrectionEquation}
    \bm\mu^E_\mathcal{A} = (\bm E \bm Q_F\bm E^T)^{-1}\bm E\bm m,
\end{align}
without approximation error and where $\bm\mu^E_\mathcal{A}$ are the parameter estimates under the Exact transformation method.
\end{proposition}
\begin{proof}
From Theorem 2.5 in \cite{rue2005gaussian}, the conditional precision matrix is $\bm Q_{B}$ for a single linear model under the constraint set $\bm A$. See Equation \eqref{eq:QPartition} for a definition of the partition of $\bm Q$. We can find the corresponding precision matrix by calculating $\bm E\bm Q\bm E^T=\bm Q_{B}$. Now $\bm E$ is defined for a single model only, and not for all models jointly. We know from Equation \eqref{eq:MeanCorrectionEquation} that the parameter estimates, including both 0-constrained values and non-zero estimates, are given by $\bm \Sigma_A^*\bm Q\bm\mu$. Instead of working with $\bm \Sigma_A^*$, we can remove the rows and columns that are 0, i.e. by considering $(\bm E \bm X^T\bm X\bm E^T)^{-1}$. We should remove the corresponding rows from $\bm Q\bm\mu$, i.e. by considering $\bm E\bm Q\bm \mu$. There is no approximation error as no constraint is enforced approximately. For the full set of constraints, we can collect the constraints in the matrix $\mathcal{\bm A}$. We pad with 0's. This matrix now contains all constraints. Due to the block diagonal structure (the block diagonal structure means the models are independent) of $\bm Q_F$, the formula for the single model generalizes to the regression with all parameters that are not constrained to be 0.   
\end{proof}

\subsection{Computational algorithms} 
\label{sec:Algorithms}
We have proposed three new methods, the Diagonal correction method, the Projection method and the Exact transformation method. We have collected all the constraints in the matrix $\mathcal{\bm A}$ and by careful construction of the joint precision matrix, we can obtain either approximate (the first two methods) or exact estimates (the last method) of the regression coefficients for all submodels of a linear model.

In Algorithm \ref{alg:DiagonalAlgorithm} we outline the code for the estimation of coefficients using the Diagonal correction method. In Algorithm \ref{alg:ProjectionAlgorithm} we provide the pseudo-code for the Projection method and in Algorithm \ref{alg:ExactAlgorithm} we provide the pseudo-code for the Exact transformation method. 

Note that we do \textit{not} calculate the inverse of e.g. $\bm X^T \bm X+ \bm A^T\bm A\kappa$, but instead calculate the Cholesky decomposition of this matrix and solve the corresponding linear system of equations. However, instead of solving each such system iteratively, we estimate all/several of the systems of equations jointly, as outlined in Algorithm \ref{alg:DiagonalAlgorithm}. This also holds for the other algorithms. We do not perform matrix inversions, but perform Cholesky factorizations for \textit{very} sparse matrices.  

If our goal is to estimate a regression model with only one feature,  we are  better off by estimating that model and not by correcting the larger model. However, if we want to fit many models, the bookkeeping of coefficients becomes more tedious and time consuming. Also, normally, the precision matrix $\bm Q$ is not too big. In space-time models the joint precision matrix with main and interaction effects $\bm Q$ can have dimensions 10 000 by 10 000. In our setting, typically $p$ is in the range between 10 and 20, so that $\bm Q$ is small in dimension, even if some of the covariates are categorical with several categories. This means a joint precision matrix, where several independent models are considered jointly, is very sparse.  
 
\begin{algorithm}
\caption{Pseudo-code for the Diagonal correction method. }\label{alg:DiagonalAlgorithm}
\textbf{Input}: Design matrix $\bm X$, which means $\bm Q=\bm X^T\bm X$, model predictions $\bm f$ using the training data, feature vector $\bm d^*$ to be explained.\\
\textbf{Output}: Estimated Shapley values $\bm\phi$ for one or several observations.

\textbf{Procedure processing}:
Calculate $\bm Q_F=\bm I_{2^p}\otimes \bm Q$, where $\bm I_{2^p}$ is the $2^p\times 2^p$ identity matrix and $\bm Q= \bm X^T \bm X$. Calculate $\mathcal{\bm A}^T\mathcal{\bm A}$, where $\mathcal{\bm A}$ is the joint constraint matrix of all submodels. 
Calculate $\bm m$ as $\bm X^T\bm f$ repeated $2^p$ times.\\
\textbf{Procedure calculations:}\\
\textbf{Compute} the Cholesky decomposition (or a generalization of it), $\bm L$, of $(\bm Q_F+\mathcal{\bm A}^T\mathcal{\bm A}\kappa)$.
The scalar value $\kappa$ is set to be e.g. $\kappa=10^5 \lambda_{max}$, i.e. the largest eigenvalue of $\bm Q$ multiplied with $10^5$.\\\ 
\textbf{Solve:} Equation \eqref{eq:DcorrectionEquation}.\\
\textbf{Obtain} $\bm \phi$ by solving Equation \eqref{eq:kernelShap} using the estimated $v$'s where we have used $\bm d^*$ (the observation to be explained) as feature vector.  
\end{algorithm}

\begin{algorithm}
\caption{Pseudo-code for the Projection method.}\label{alg:ProjectionAlgorithm}
\textbf{Input}: Design matrix $\bm X$, which means $\bm Q=\bm X^T\bm X$, model predictions $\bm f$ using the training data, feature vector $\bm d^*$ to be explained.\\
\textbf{Output}: Estimated Shapley values $\bm\phi$ for one or several observations.\\
\textbf{Procedure processing}:
Calculate $\bm Q_F=\bm I_{2^p}\otimes \bm Q$, where $\bm I_{2^p}$ is the $2^p\times 2^p$ identity matrix and $\bm Q= \bm X^T \bm X$. Calculate $\bm Z=\bm I-\mathcal{\bm A}^T\mathcal{\bm A}$, where $\mathcal{A}$ is the joint constraint matrix of all submodels. 
Calculate $\bm m$ as $\bm X^T\bm f$ repeated $2^p$ times.\\
\textbf{Procedure calculations:}\\
\textbf{Compute} the Cholesky decomposition (or a generalization of it), $\bm L$, of $(\bm Z\bm Q_F\bm Z+\varepsilon\bm I)$.
The scalar value $\varepsilon$ is set to be e.g. $\varepsilon=10^{-5}$. \\\ 
\textbf{Solve:} Equation \eqref{eq:PcorrectionEquation}.\\
\textbf{Obtain} $\bm \phi$ by solving Equation \eqref{eq:kernelShap} using the estimated $v$'s where we have used $\bm d^*$ (the observation to be explained) as feature vector.  
\end{algorithm}

\begin{algorithm}
\caption{Pseudo-code for the Exact transformation method. }\label{alg:ExactAlgorithm}
\textbf{Input}: Design matrix $\bm X$, which means $\bm Q=\bm X^T\bm X$, model predictions $\bm f$ using the training data, feature vector $\bm d^*$ to be explained.\\
\textbf{Output}: Estimated Shapley values $\bm\phi$ for one or several observations.\\
\textbf{Procedure processing}:
Calculate $\bm Q_F=\bm I_{2^p}\otimes \bm Q$, where $\bm I_{2^p}$ is the $2^p\times 2^p$ identity matrix and $\bm Q= \bm X^T \bm X$. Compute the mapping matrix $\bm E$.   
Calculate $\bm m$ as $\bm X^T\bm f$ repeated $2^p$ times.\\
\textbf{Procedure calculations:}\\
\textbf{Compute} the Cholesky decomposition (or a generalization of it), $\bm L$, of $\bm E\bm Q_F\bm E^T$.
\textbf{Solve:} Equation \eqref{eq:EcorrectionEquation}.\\
\textbf{Obtain} $\bm \phi$ by solving Equation \eqref{eq:kernelShap} using the estimated $v$'s where we have used $\bm d^*$ (the observation to be explained) as feature vector.  
\end{algorithm}

We implement the code in a R-script \citep{R_language} using native R functions and functions mainly from the Matrix \citep{MatrixBates} package for sparse matrix algebra. We use default BLAS (Basic Linear Algebra Subprograms), so that we do not perform any parallel computations with our methods. In comparison, shapr uses the future package to obtain parallelism. We also re-use some R-functions already implemented in \cite{kernelshap}. We do not make a new R-package suited for 
constructing the matrices. 

\subsection{Brief summary}
In the three new methods, we construct a global model with a sparse precision matrix. All three methods exploit the high proportion of zeros in the joint precision matrices. Most zeros are not stored explicitly, and an off-diagonal zero in the precision matrix indicates conditional independence. Due to the block diagonal structure in each of the joint precision matrices, we obtain independent models, so that parameter estimates from one model do not affect the estimates in other models. When plotting such a matrix, the non-zero elements are close to the main diagonal. For the two approximate methods, there are tuning parameters involved. They should be chosen appropriately.        

\section{Numerical case studies}\label{sec:numericalSection}
We now perform numerical studies. We run the experiments using a machine with a processor with specifications 12th Gen Intel(R) Core(TM) i9-12900H   2.50 GHz, 64 GB RAM and a Windows operating system. We study the Adult dataset \citep{adult_2}, a simulated dataset and the WHO Life Expectancy dataset from Kaggle. Often, the dataset is split into train and test sets. In \cite{olsen2023comparative} they split the Adult dataset into a training (30,000) and a test (162) data set. Shapley values are estimated for the test data only.  With this choice, we explain very few observations. Also, our methods do not actually see the value of the response variable of the \textit{new} observations we want to explain. We use all data for training and to make explanations. We can pretend that we have new realizations of the response variable for the same predictor variables when we explain the training data.      

\subsection{Adult income dataset}
\label{sec:AdultMainNumerical}
We consider a subset of 11 covariates from the original Adult dataset and the outcome variable. We consider Work-class (categorical), fnlwgt (continuous), Education (categorical), Marital status (categorical), Occupation (categorical),  Relationship status (categorical), Race (categorical), Gender (categorical), Capital gain (continuous), Capital loss (continuous)  and Hours per week (continuous). The prediction task is to determine whether a person makes over 50K a year. We fit a linear model to the predicted probabilities obtained using a logistic regression with logit link function. We use a linear explainer to explain this linear model. We use the 30 162 observations both for training and for obtaining explanations. We find the largest eigenvalue $\lambda_{max}$ of $\bm Q$ and set $\kappa =\lambda_{max}\cdot 10^{5}$, which is the default suggestion. We use $\varepsilon =10^{-5}$. 

The Iterative method in shapr takes about 17 minutes to run and uses all coalitions. The Sequential method runs in 19 minutes. The Diagonal correction method takes 2.5 seconds to run, the Exact transformation method takes 3.7 seconds to run and the Projection method takes 9 seconds to run.

\begin{figure}[htbp]
    \centering
    \begin{subfigure}[b]{0.48\textwidth}
        \centering
        \includegraphics[width=\linewidth]{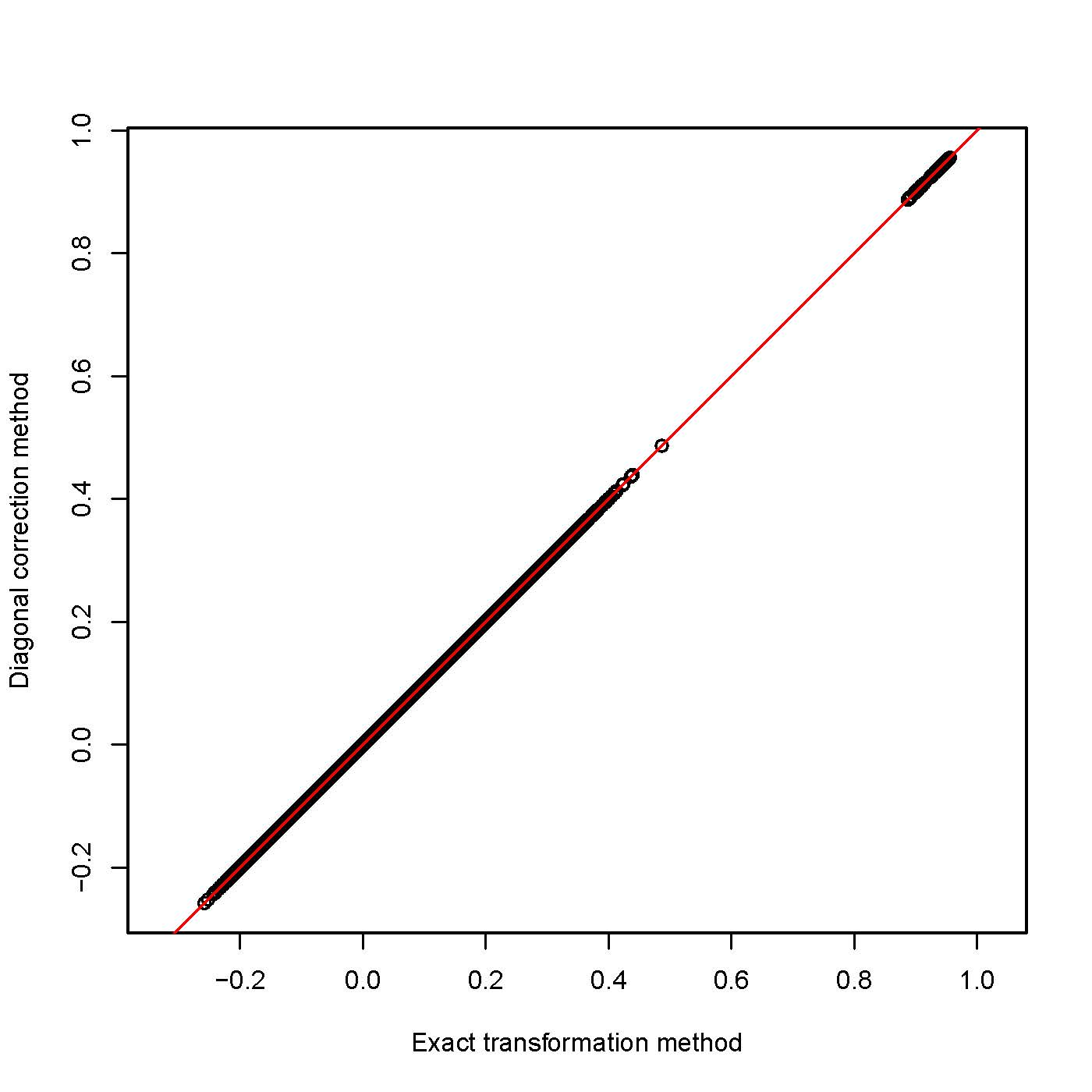}
        \caption{Along the x-axis we find the estimates
from the Exact transformation method, while along the y-axis we find the estimates from
the Diagonal correction method. }
        \label{fig:adult_E_D}
    \end{subfigure}%
    \hfill
    \begin{subfigure}[b]{0.48\textwidth}
        \centering
        \includegraphics[width=\linewidth]{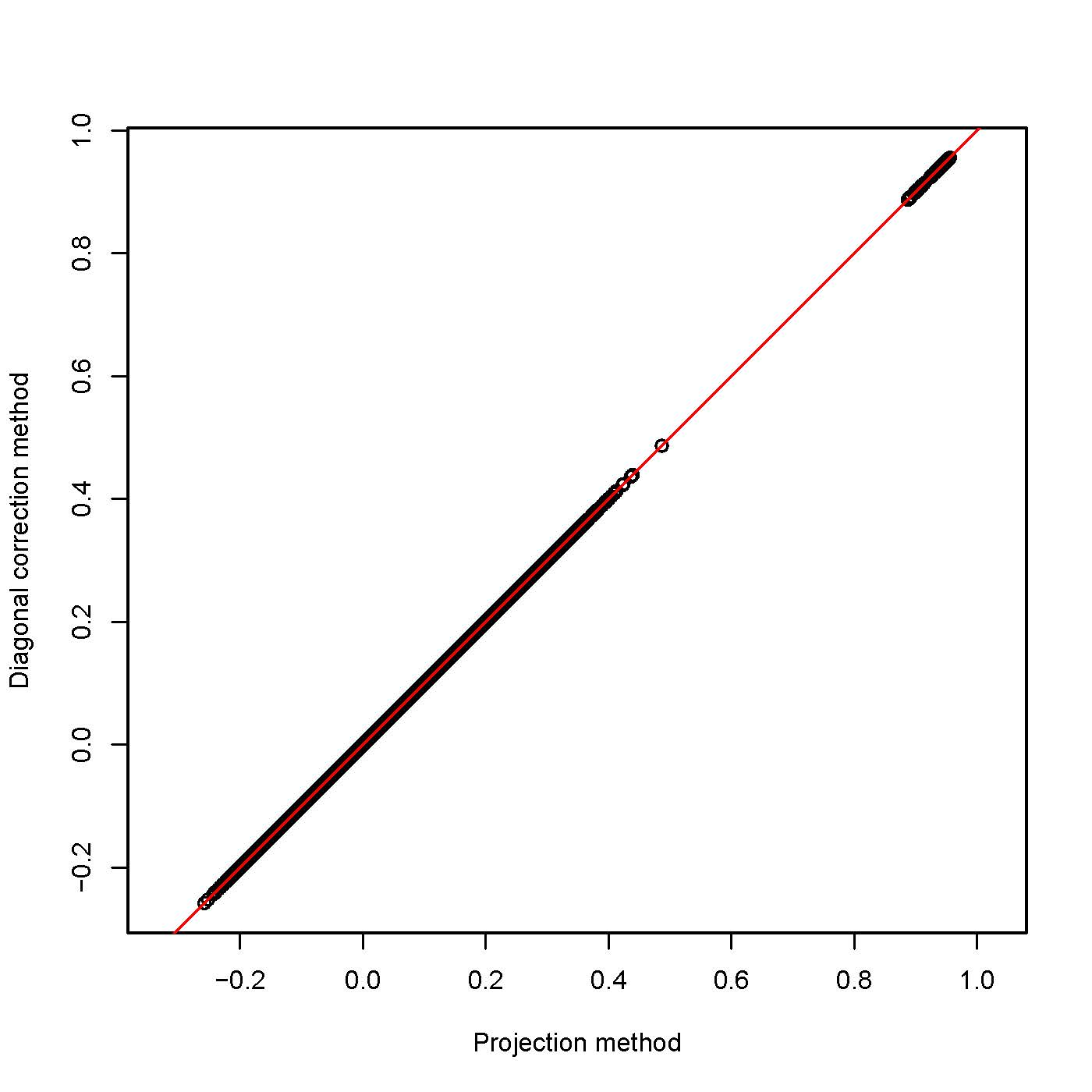}
        \caption{We plot the estimates from the Projection
method along the x-axis, while along the y-axis we plot the estimates from the Diagonal
correction method. }
        \label{fig:Adult_P_D}
    \end{subfigure}

    \medskip

    \begin{subfigure}[b]{0.48\textwidth}
        \centering
        \includegraphics[width=\linewidth]{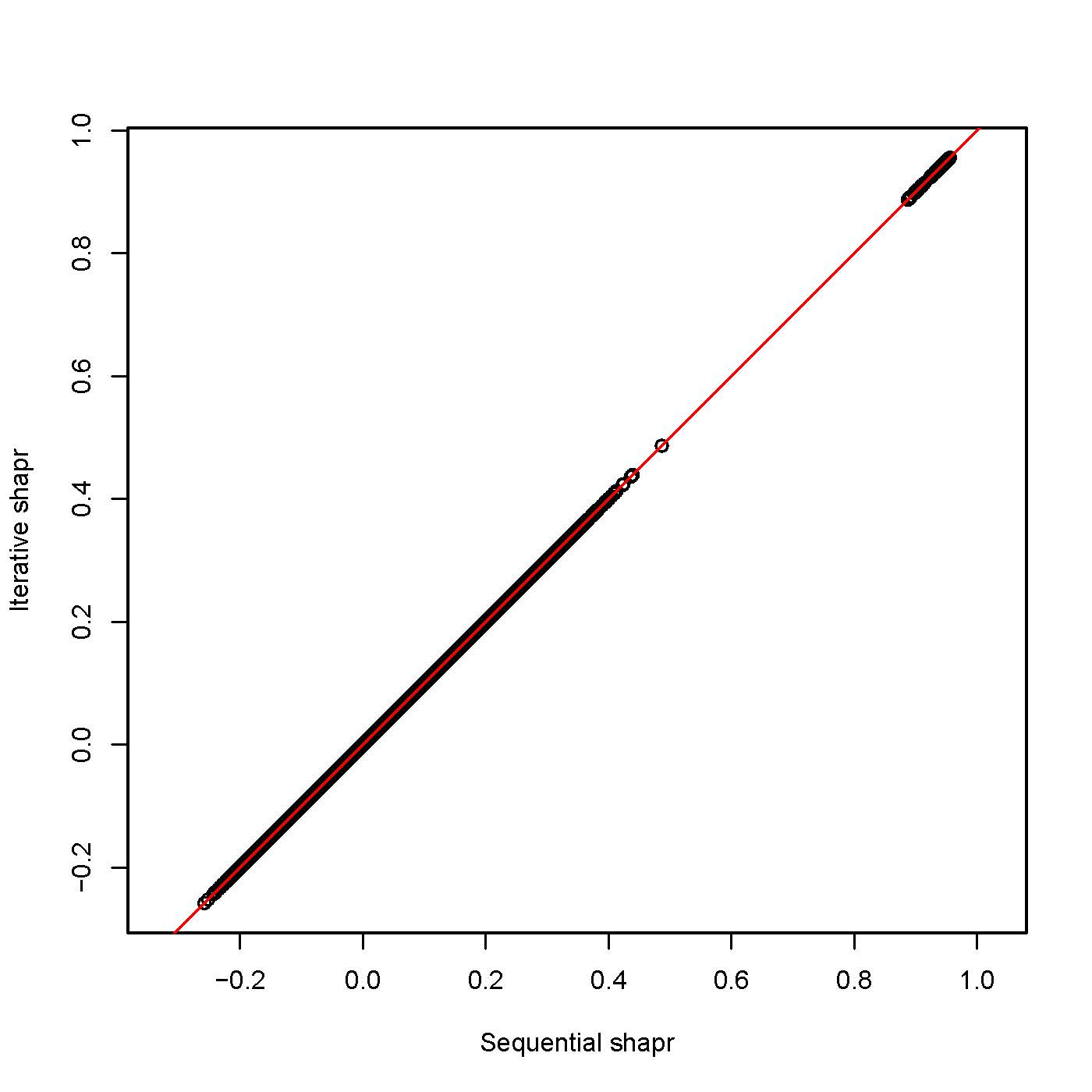}
        \caption{We plot the estimates from the shapr
sequential method along the x-axis, while along the y-axis we plot the estimates from the
shapr iterative method. }
        \label{fig:Adult_iterative_sequential}
    \end{subfigure}
    \caption{Adult dataset: Shapley value estimates from the different methods are compared. The red line $y=x$ is also shown.}
    \label{fig:Adult_comparison_methods}
\end{figure}

In Figure \ref{fig:adult_E_D}
 we plot the Shapley value estimates from the Diagonal correction method against the estimates from the Exact transformation method. In Figure \ref{fig:Adult_P_D} we compare the Shapley value estimates from the Diagonal correction method with the estimates from the Projection method. We plot all Shapley values in one plot. We observe excellent correspondence in both plots. In Figure \ref{fig:Adult_iterative_sequential} the estimates from the shapr iterative method and the shapr sequential method are plotted against each other in one plot. The correspondence is excellent. 
 
\begin{figure}
\centering
\includegraphics[width=0.95\textwidth]{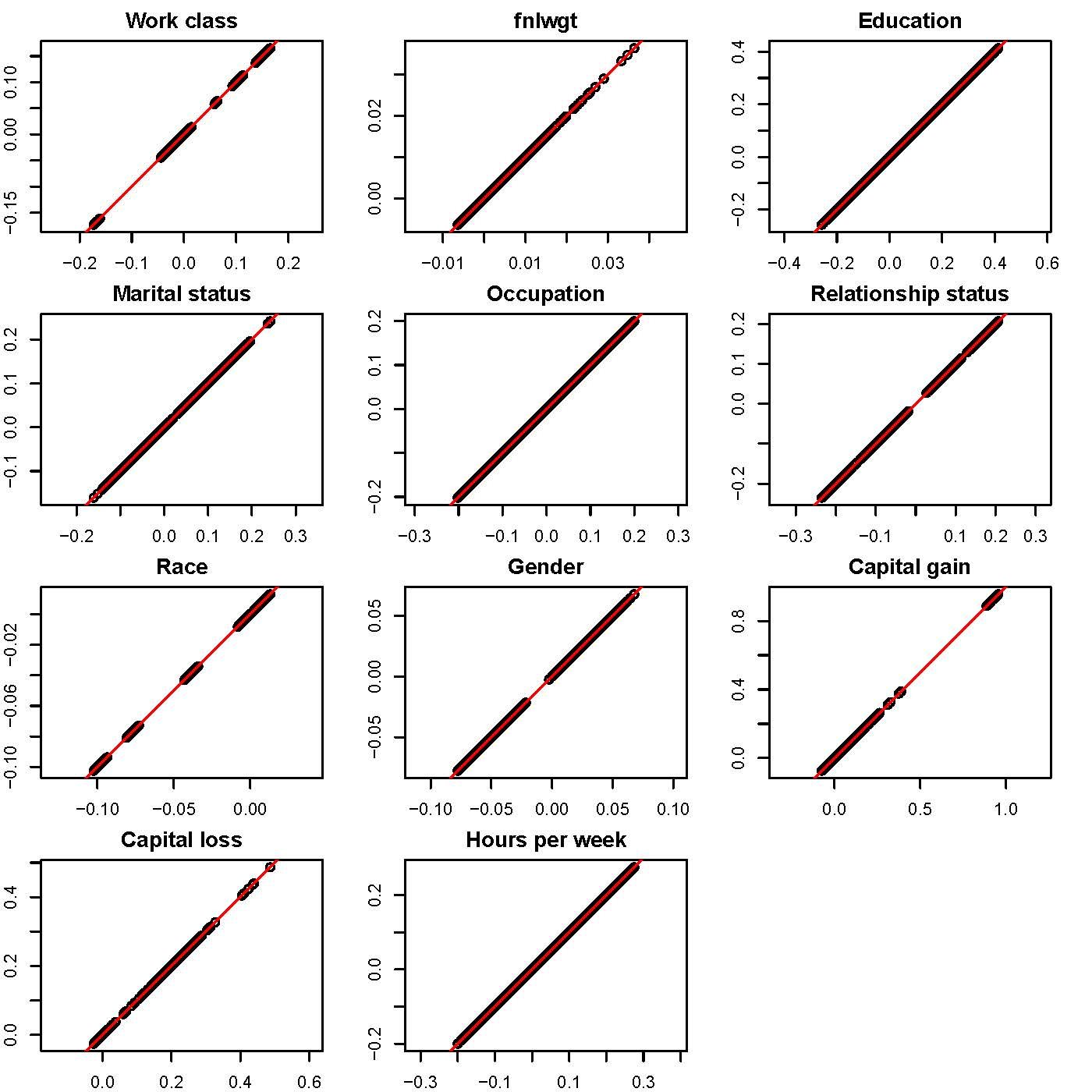}
\caption{Adult dataset: We plot the estimates of the Shapley values from the Diagonal correction method along the x-axis, while along the y-axis we plot the estimates from the shapr iterative method. The red line $y=x$ is also shown in each plot.}
\label{fig:Adult_iterative_D}
\end{figure}

 In Figure \ref{fig:Adult_iterative_D} we compare the estimated Shapley values from the shapr iterative method variable by variable against the estimates of the Diagonal correction method. The correspondence is excellent. 

 In general, all methods studied are able to estimate the Shapley values with excellent accuracy. However, the newly suggested algorithms are \textit{much} faster. The Iterative method in shapr takes almost as long to run as the Sequential method. The reason is that all coalitions are being used. This can be verified by studying the contents of the $iter\_info\_dt$ element of the $iterative\_results$ list element of the output-object from shapr.  

In Table \ref{tab:adultMAE} we show the median absolute error between the Shapley value estimates from the sequential method and the estimates from other methods. We observe that the median absolute errors are very small for all methods, indicating very good correspondence between estimates for each method. 

\begin{table}[ht]
\centering
\begin{tabular}{|l |l|}
\hline
\textbf{Method} & \textbf{MedAE} \\ \hline
Exact transformation method           &  $9.68\cdot 10^{-9}$ \\              Projection method           &  $9.74\cdot 10^{-9}$\\ Diagonal correction method  &  $9.93\cdot 10^{-9}$\\
shapr iterative method & 0\\
\hline

\end{tabular}\caption{Adult dataset: The median absolute error between the Shapley values from the Sequential shapr method and each comparison method are calculated and shown. The value 0 indicates that the iterative method uses all coalitions.}
\label{tab:adultMAE}
\end{table}

\subsection{Simulated dataset with 21 Gaussian predictors}
\label{sec:SimulationStudy}
We make a simulated dataset with 22 Gaussian distributed random variables (one response variable and the remaining 21 variables are predictor variables). Each has a marginal variance of 1 and a correlation between 0.2 and 0.3 with the other random variables (drawn uniformly between 0.2 and 0.3). We only simulate the lower triangular part of the covariance matrix and use the forcesymmetric-matrix-function with the uplo-argument set to ``L'' to create the full covariance matrix. The reason we pick 22 variables is due to memory constraints on our machine. We name the variables according to the alphabet and number system so that the first covariate is named a1, the second one b2 and so on up to u21. The first variable in the dataset is named Y and is the response variable. We create 1000 realizations and explain the 100 first. We apply the linear regression model using the $Y$-variable as outcome variable. We explain the predictions using the linear explainer. We find the largest eigenvalue $\lambda_{max}$ of $\bm Q$ and set $\kappa =\lambda_{max}\cdot 10^{5}$, which is the default suggestion. We use $\varepsilon =10^{-2}$ as the smallest eigenvalue of $\bm Q$ is 732.38. As we will see, the numerical results are still very good for the Projection method. This shows that other choices of $\varepsilon$ are possible than the one suggested.  

The Diagonal correction method takes 1.4 minutes to run. The Exact transformation method takes about 2.2 minutes to run. The Projection method takes about 2.5 minutes to run. The Iterative method in shapr runs in 6.48 seconds and uses 200 coalitions. Due to its cost, we do not run the Sequential method in the shapr package.  

\begin{figure}[htbp]
    \centering
    \begin{subfigure}[b]{0.48\textwidth}
        \centering
        \includegraphics[width=\linewidth]{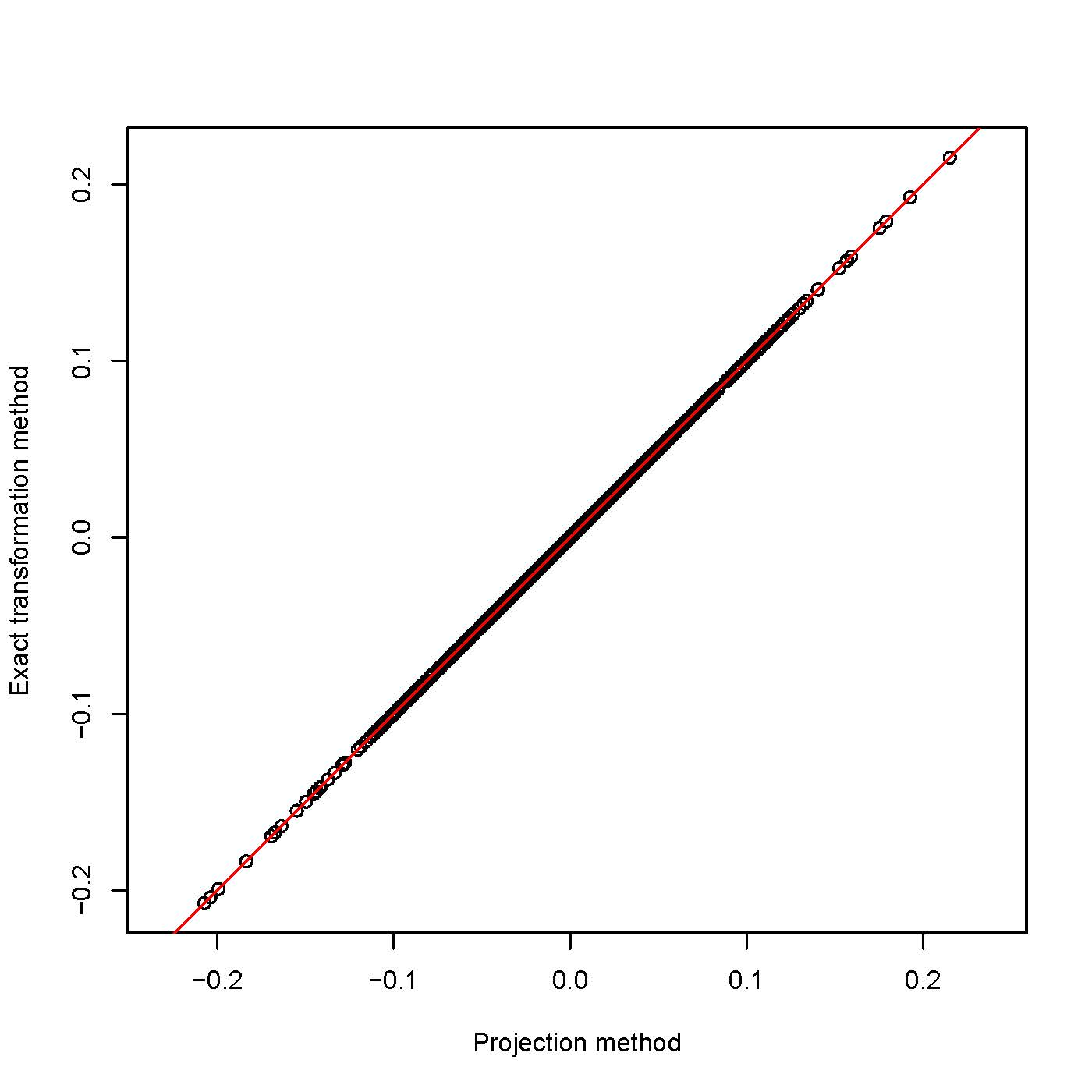}
        \caption{We show along the x-axis estimates from the Projection method, while along the y-axis are estimates from the Exact transformation method.}
        \label{fig:Simulationstudy_P_vs_E}
    \end{subfigure}%
    \hfill
    \begin{subfigure}[b]{0.48\textwidth}
        \centering
        \includegraphics[width=\linewidth]{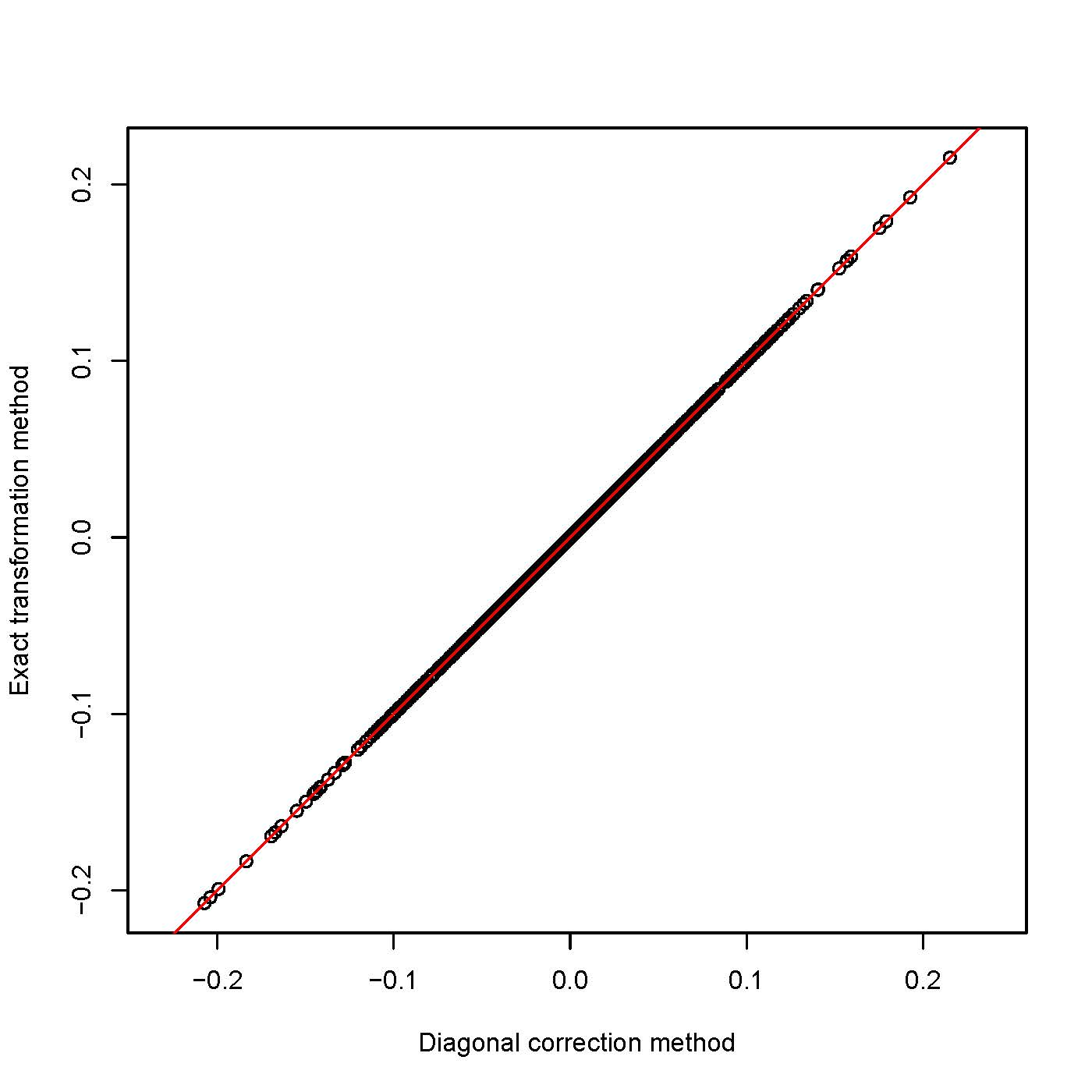}
        \caption{We show along the x-axis estimates from the Diagonal correction method, while along the y-axis are estimates from the Exact transformation method.  }
        \label{fig:Simulationstudy_D_vs_E}
    \end{subfigure}
    \caption{Simulated data: We compare the Shapley value estimates between the new methods. The red line $y=x$ is plotted in each subfigure.}
    \label{fig:Adult_comparison_methods}
\end{figure}

In Figure \ref{fig:Simulationstudy_P_vs_E} we compare the Shapley value estimates from the Exact method against the estimates from the Projection method. In Figure \ref{fig:Simulationstudy_D_vs_E} we compare the Exact transformation method against the Diagonal correction method. In both figures there is excellent correspondence.  

\begin{figure}
\centering
\includegraphics[width=0.65\textwidth,trim=0.05cm 0.55cm 0 0.55cm, clip]{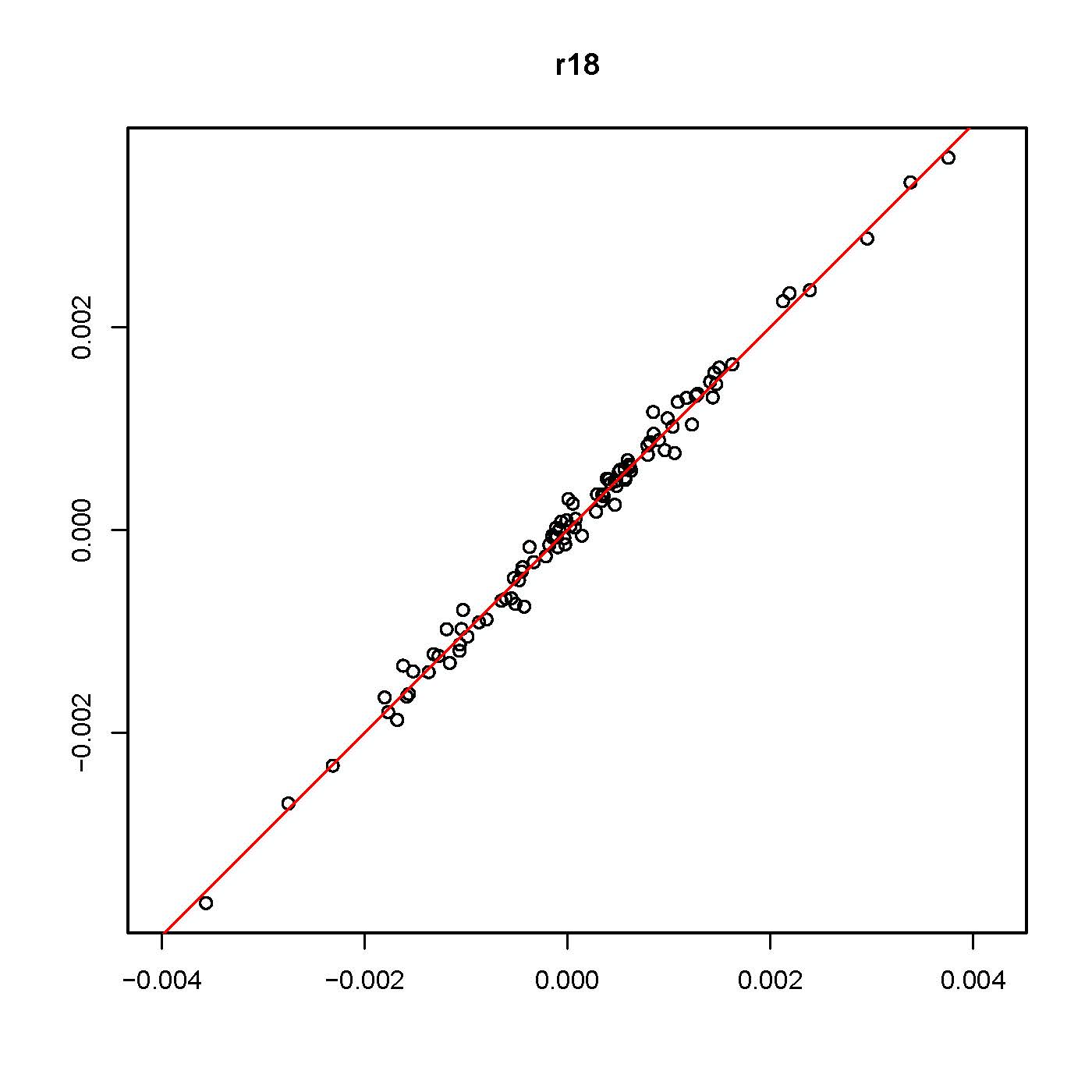}
\caption{Simulated data: The Shapley value estimates for the variable r18. Along the x-axis we find the estimates from the Diagonal correction method, while along the y-axis we find the estimates from the shapr iterative method.}
\label{fig:Simulationstudy_r18}
\end{figure}

\begin{figure}
\centering
\includegraphics[width=0.95\textwidth,trim=0.0cm 0.00cm 0 0.00cm, clip]{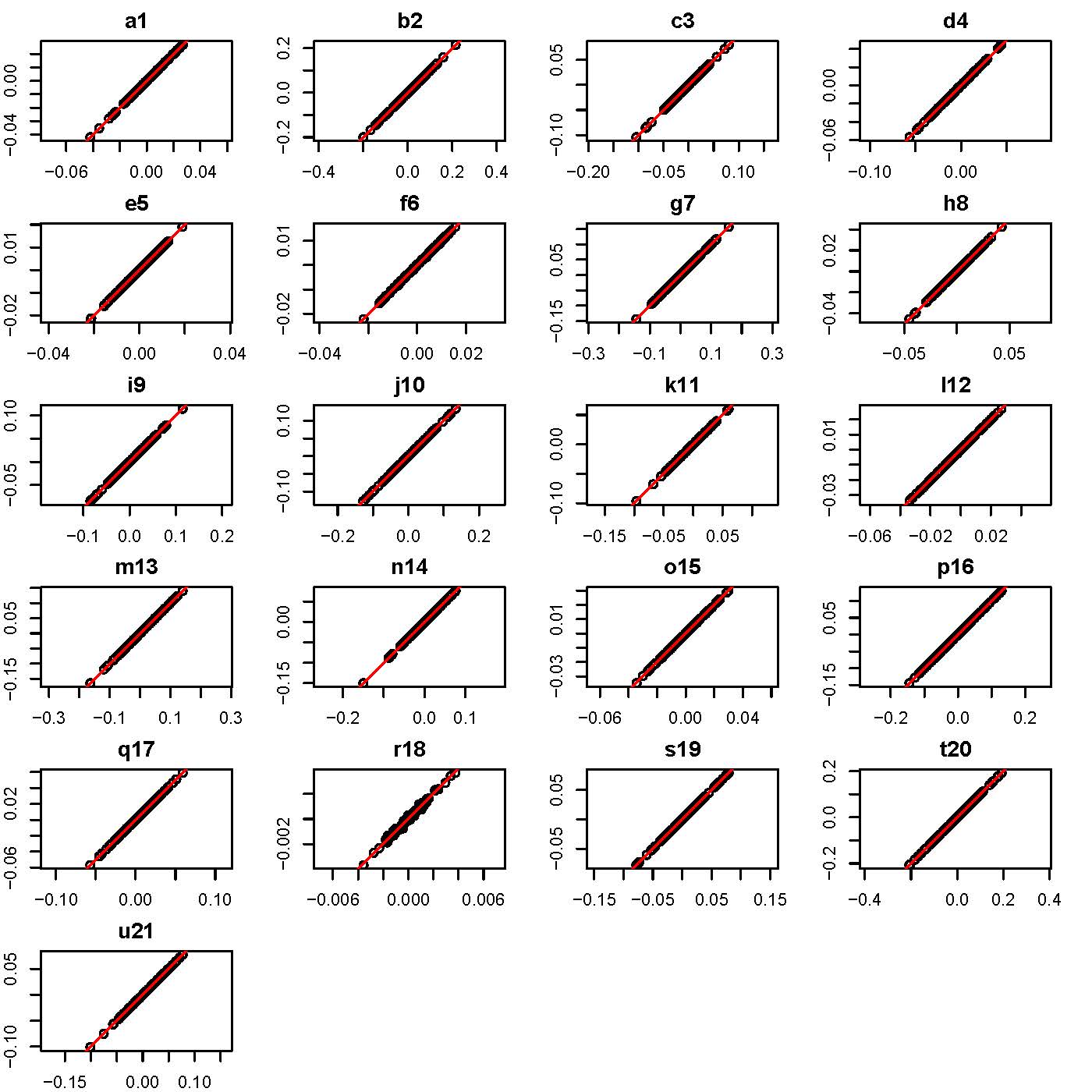}
\caption{Simulated data: The estimated Shapley values for each variable are shown. We plot along the x-axis the estimates from the Diagonal correction method, while along the y-axis we plot the estimates from the shapr iterative method. The red line $y=x$ is shown in each subplot.}
\label{fig:Simulationstudy_iterative_D}
\end{figure}

In Figure \ref{fig:Simulationstudy_r18} we plot the variable with worst correspondence between the Iterative method of shapr and the new algorithms, namely r18. The correspondence is good, but not excellent. In Figure \ref{fig:Simulationstudy_iterative_D} we plot the estimated Shapley values from the Iterative shapr method with the estimates from the Diagonal correction method. The correspondence is excellent, except for r18.

In this case study the Iterative method of shapr is much faster, with comparable accuracy to the newly proposed methods. However, our methods estimate 2 097 152 models, while shapr estimates 200 submodels. We note that it takes only 2.5 minutes or less to estimate all model coefficients for the new methods. 

In Table \ref{tab:Simulated_MAE} the calculated median absolute errors are very low, indicating very good correspondence between the methods.

\begin{table}[ht]
\centering
\begin{tabular}{|l |l|}
\hline
\textbf{Method} & \textbf{MedAE} \\ \hline         Projection method           &  $1.88\cdot 10^{-7}$\\ Diagonal correction method  &  $ 1.21\cdot 10^{-7}$\\
shapr iterative method &  $8.32\cdot 10^{-5}$\\
\hline

\end{tabular}\caption{Simulated dataset: The median absolute error between the Shapley values from the Exact transformation method and each comparison method are calculated and shown. Note: we have not used the Sequential method in shapr due to the large number of submodels.}
\label{tab:Simulated_MAE}
\end{table}

\subsection{WHO dataset}

We use the Life Expectancy (WHO) Fixed dataset from Kaggle. We only use a subset of the covariates, to be more precise 16 of them, so that $\bm Q$ is full rank. In Table \ref{tab:WHOvariablenames} we provide an overview of the variables studied. We use the variable Life expectancy as response variable in a linear regression. We apply the linear explainer on the predictions from this model. There are 2864 observations and we use all observations for training and also for obtaining explanations. We find the largest eigenvalue $\lambda_{max}$ of $\bm Q$ and set $\kappa =\lambda_{max}\cdot 10^{5}$, which is the default suggestion. We use $\varepsilon =10^{-5}$. 

\begin{table}[h]
\centering
\caption{Overview of variables considered in the WHO dataset}
\label{tab:WHOvariablenames}
\begin{tabular}{l l}
\hline
\textbf{Short Variable Name} & \textbf{Description} \\ \hline
Life Expectancy    & Expected age of death in years    \\ 
Country    & Country    \\ 
Year & Year \\
Infant deaths & Number of infant deaths per 1 000.\\
Under five deaths & Number of under-five-year-old-deaths per 1 000\\
Adult mortality & The probability of dying between 15 and 60 years \\
Alcohol consumption & Alcohol consumption in liters of pure alcohol per capita \\
Hepatitis B & Hepatitis B immunization coverage among 1-year-olds\\
Measles & Number of recorded cases of measles per 1000  \\
BMI & Average Body Mass Index of entire population\\
Polio & Immunization coverage among 1-year-olds\\
Diphtheria & Diphtheria tetanus toxoid and pertussis immunization coverage\\
Incidents HIV &  Deaths per 1 000 live births\\
GDP per capita & Gross Domestic Product per capita in USD\\
Thinness ten to nineteen & Prevalence of thinness among population between ten and nineteen\\
Thinness five to nine &  Prevalence of thinness among children between five and nine\\
Schooling & Number of years with education\\ \hline
\end{tabular}
\end{table}

The Sequential method takes 131 minutes to run, which is about 2.2 hours. The method uses the future package for calculating the contribution functions in parallel. The Iterative method in parallel mode takes about 16 minutes to run. It uses 2 744 coalitions out of 65 536. In serial mode it takes about 36 minutes to run. 

The Projection method takes about 1.2 minutes to run. The Diagonal correction method also takes about 1.2 minutes to run. The Exact transformation method runs in about 1.1 minutes. 

\begin{figure}[htbp]
    \centering
    \begin{subfigure}[b]{0.48\textwidth}
        \centering
        \includegraphics[width=\linewidth]{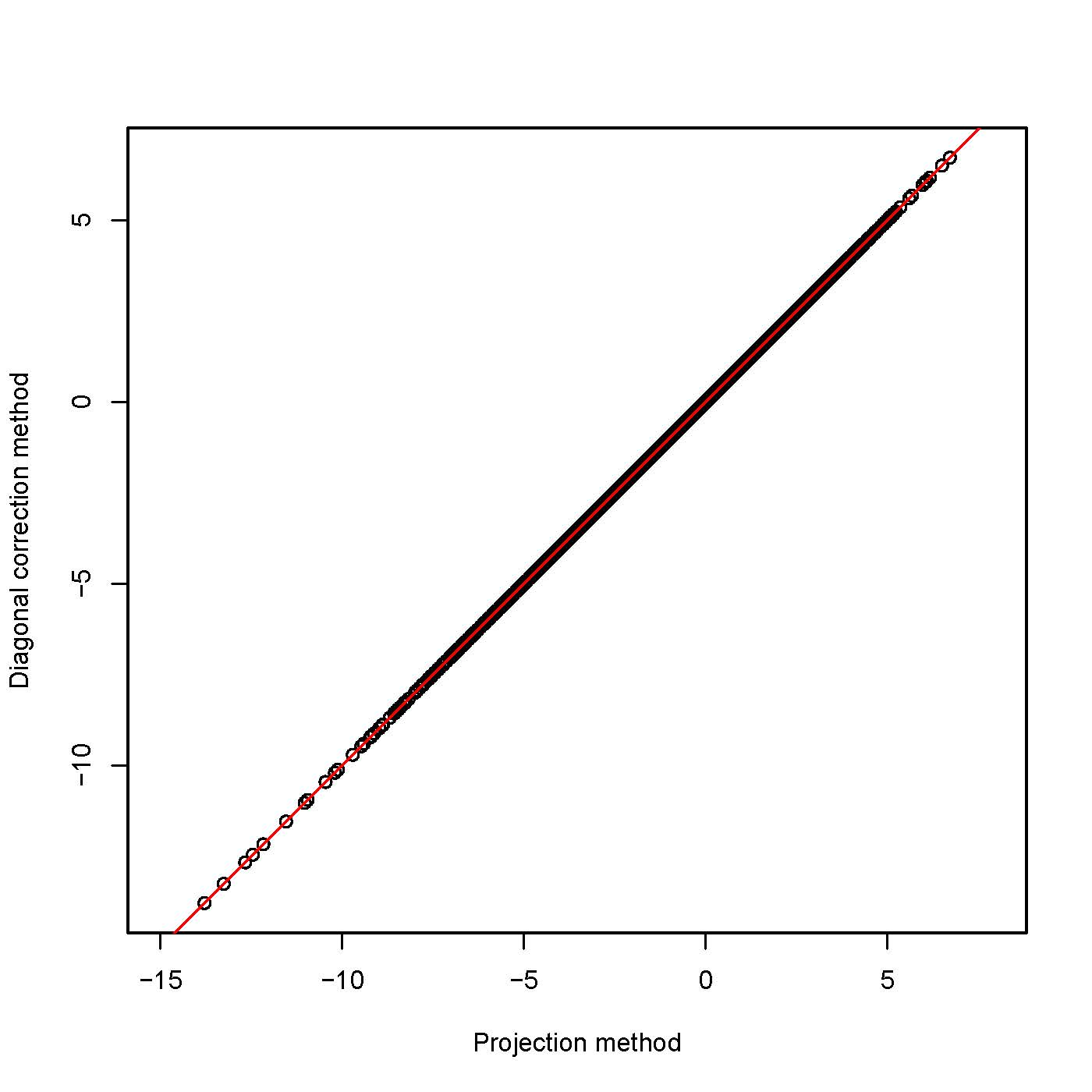}
        \caption{Along the x-axis are the estimates from the Projection method, while along the y-axis are the estimates from the Diagonal correction method.}
        \label{fig:WHO_P_D}
    \end{subfigure}%
    \hfill
    \begin{subfigure}[b]{0.48\textwidth}
        \centering
        \includegraphics[width=\linewidth]{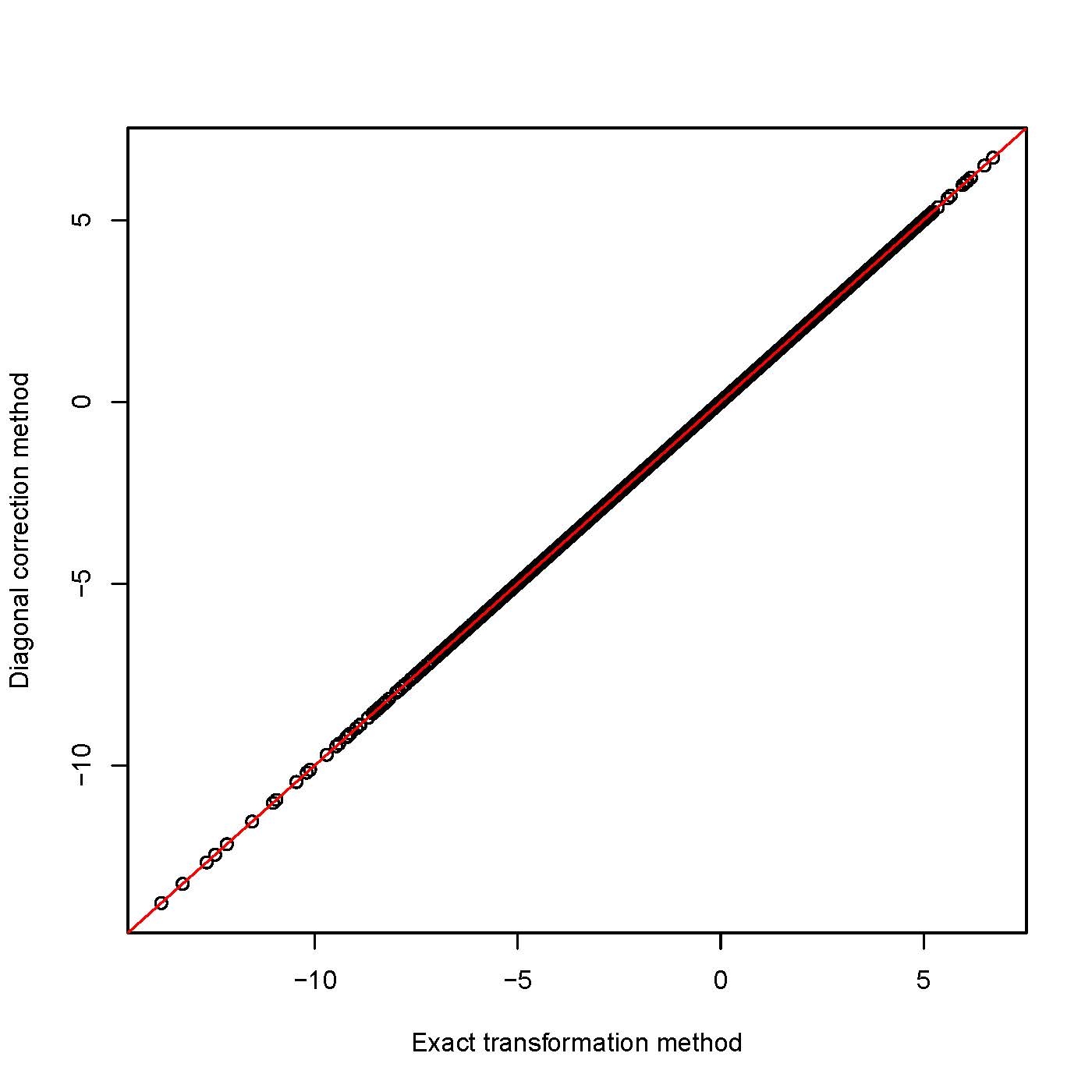}
        \caption{ Along the x-axis we plot the output from the Exact transformation method, while along the y-axis we plot the output from the Diagonal correction method.}
\label{fig:WHO_E_D}
    \end{subfigure}
        \medskip

    \begin{subfigure}[b]{0.48\textwidth}
        \centering
        \includegraphics[width=\linewidth]{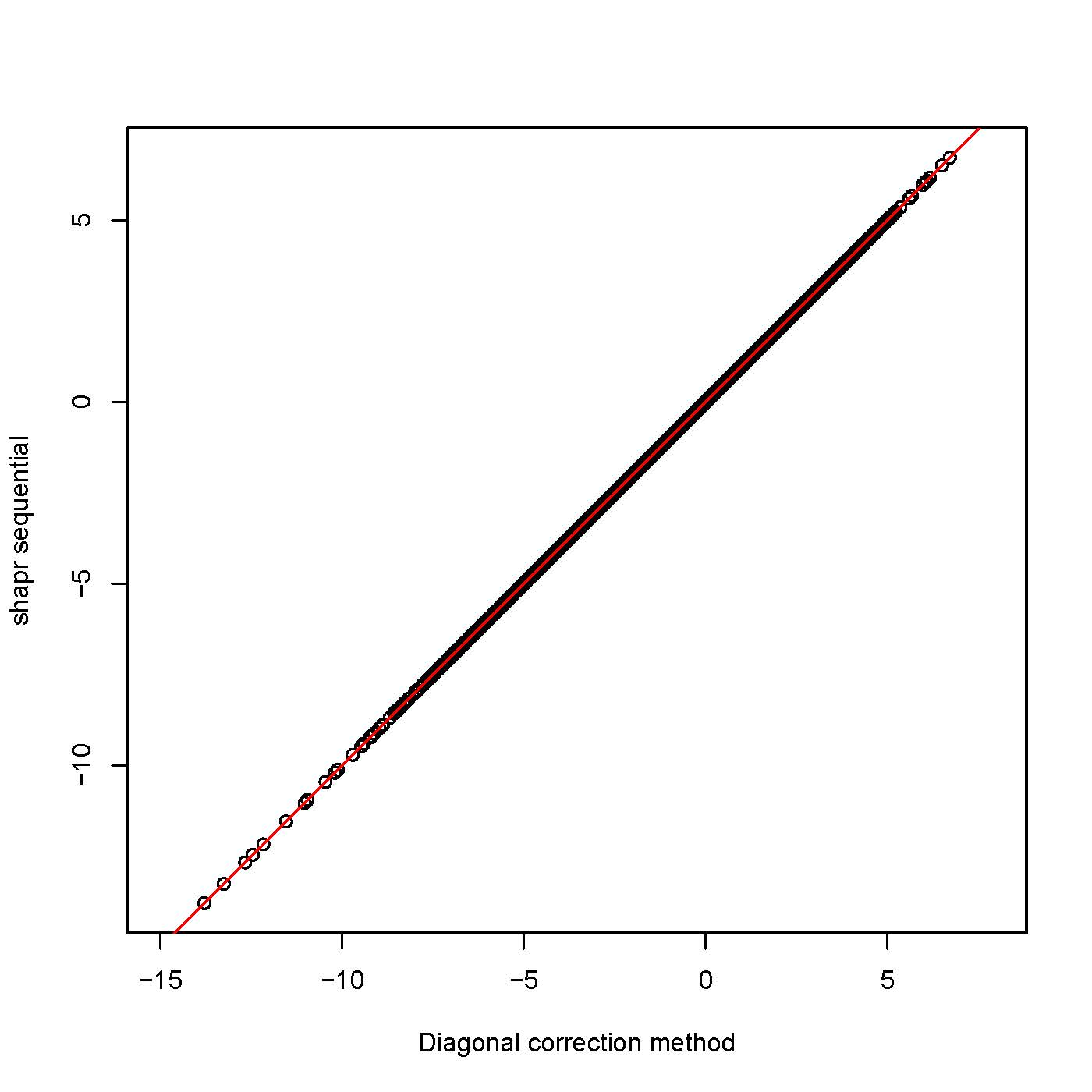}
        \caption{Along the x-axis we plot the output from the Diagonal correction method, while along the y-axis we plot the estimates from the Sequential method in shapr.}
        \label{fig:WHO_shapr_seq_vs_diagonal}
    \end{subfigure}

    \caption{WHO dataset: Comparison of Shapley value estimates between the different methods. The red line $y=x$ is shown in each subfigure.}
    \label{fig:Adult_comparison_methods}
\end{figure}

In Figure \ref{fig:WHO_P_D} we plot the estimated Shapley values using the Projection method and the Diagonal correction method. We observe perfect correspondence. In Figure \ref{fig:WHO_E_D} we plot the Shapley values from the Diagonal correction method and the Exact transformation method.  Again, there is perfect correspondence. In Figure \ref{fig:WHO_shapr_seq_vs_diagonal} we plot the Diagonal correction method estimates against the estimates from the Sequential method in shapr. The correspondence is excellent.

\begin{figure}
\centering
\includegraphics[width=0.65\textwidth,trim=0.05cm 0.55cm 0 0.55cm, clip]{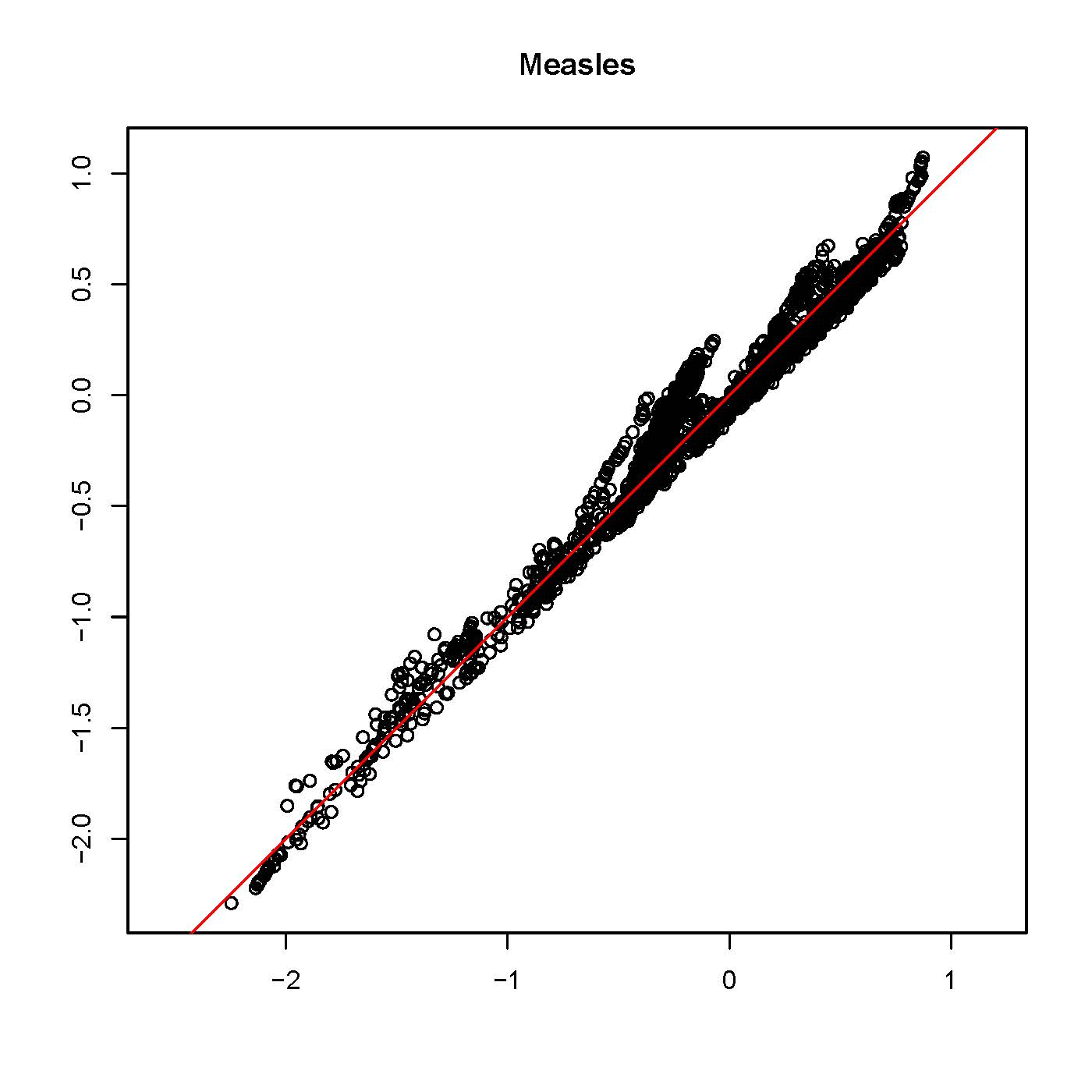}
\caption{WHO dataset: We plot the Shapley values for the Measles variable only. Along the x-axis we plot the estimates from the Diagonal correction method, while along the y-axis we plot estimates from the Iterative method in shapr. The red line along $y=x$ is also plotted.}
\label{fig:WHO_shapr_Diagonal_measles}
\end{figure}

\begin{figure}
\centering
\includegraphics[width=0.9\textwidth]{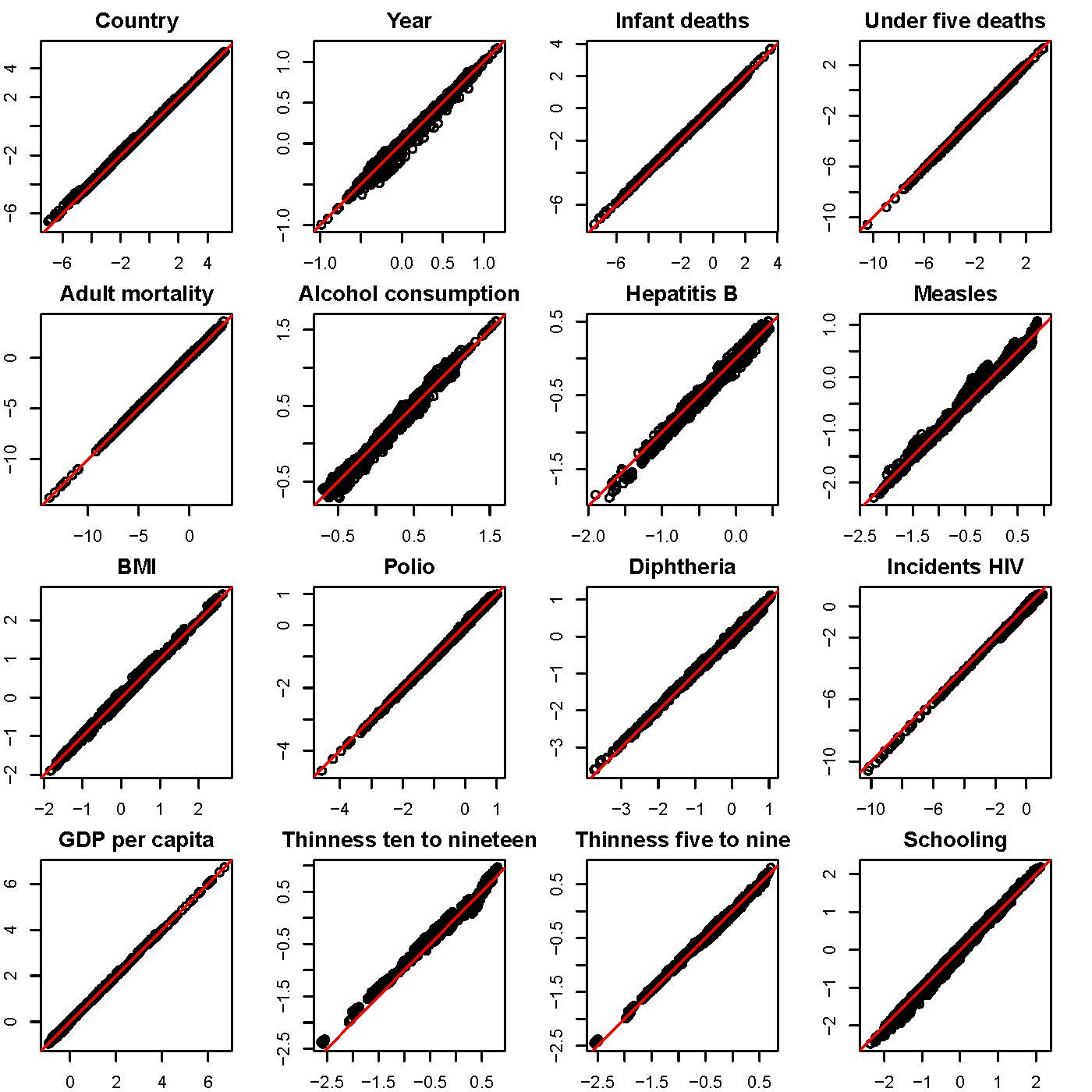}
\caption{WHO dataset: Along the x-axis we plot the output from the Diagonal correction method, while along the y-axis we plot the estimates from the shapr iterative method. A red line along $y=x$ is also plotted.}
\label{fig:WHO_shapr_iterative_vs_D}
\end{figure}

In Figure \ref{fig:WHO_shapr_Diagonal_measles} we plot one of the variables where the correspondence is worst between the Iterative shapr method and the newly suggested methods.  And in Figure \ref{fig:WHO_shapr_iterative_vs_D} we plot the Shapley estimates for each variable from the Iterative method against the Diagonal correction method. The correspondence is quite good, but a little more noisy than for the Simulated data case study and the Adult dataset case study. 

In Table \ref{tab:WHO_MAE} we compute the median absolute error between the Shapley value estimates from the shapr sequential method and the other methods. For the shapr sequential methods and the new methods the median absolute errors are very low, but the error when comparing the Sequential method with the Iterative method in shapr is not equally good.  

\begin{table}[ht]
\centering
\begin{tabular}{|l |l|}
\hline
\textbf{Method} & \textbf{MedAE} \\ \hline
Exact transformation method           &  $2.30\cdot 10^{-6}$ \\              Projection method           &  $3.66\cdot 10^{-6}$\\ Diagonal correction method  &  $3.06\cdot 10^{-6}$\\
shapr iterative method & 0.038\\
\hline

\end{tabular}\caption{WHO Life Expectancy dataset: The median absolute error between the Shapley values from the Sequential shapr method and each other method are calculated and shown.}
\label{tab:WHO_MAE}
\end{table}

\section{Discussion and conclusion}\label{sec:DiscussionConclusion} In Table \ref{tab:all_computing_times} we have collected all computing times for all case studies. The run-times show that the new methods estimate all $2^p$ coalitions in minutes or seconds. When the iterative method requires all or a large fraction of the coalitions, as in the Adult and WHO case studies, the new methods are substantially faster at equal or better accuracy. When few coalitions suffice, as in the simulated study, the Iterative method is faster, but the new methods still deliver full enumeration of all submodels within minutes.

There are also other competing methods not studied here, such as a method in the popular shap package \citep{Lundberg2017}, which assumes the variables follow a multivariate Gaussian distribution and also uses sampling. This is only one of several methods available. As we have categorical variables in our case studies, the assumption is violated. 

Note that the method can easily be extended to accommodate some types of splines, e.g. splines from the rms-package \citep{RMS}. The reason is that the coefficients can be estimated using a linear regression model. This makes our methodology more flexible. This is suggested future work. 

The shapr package performs parallelization. Further improvement of the new methods is possible by using parallelization. A recent method with parallel decompositions of a symmetric positive definite matrix is \cite{fattah2025stilesacceleratedcomputationalframework}, where also existing sparse factorization packages are compared to each other and the new method. We note that the parallelization makes the computing time comparison somewhat unfair for the new methods. We observe that the iterative method takes about 36 minutes in serial mode and not 16 minutes, as in the parallel mode, for the WHO dataset. For our work, it should be possible to use OpenMP and Intel MKL without making any code changes. This is suggested future work.

\begin{table}[ht]
\centering
\begin{tabular}{|l l l l|}
\hline
\textbf{Dataset}  & \textbf{Method} & \textbf{Serial or parallel}& \textbf{Computing time}\\ \hline
 Adult & Exact transformation method            & Serial &  3.7 s \\      Adult &  Projection method           & Serial & 9 s\\ Adult & Diagonal correction method  &Serial &  2.5 s\\
Adult & shapr iterative method & Parallel & 17 min \\
Adult & shapr sequential method & Parallel & 19 min \\

Simulated & Exact transformation method  & Serial          & 2.2 min\\      Simulated &  Projection method  & Serial         &  2.5 min\\ Simulated & Diagonal correction method  &  Serial & 1.4 min\\Simulated & shapr iterative method & Parallel & 6.48 s\\
Simulated & shapr iterative method & Serial & 6.56 s\\

WHO & Exact transformation method      & Serial &  1.1 min \\      WHO &  Projection method           & Serial & 1.2 min \\ WHO & Diagonal correction method  &  Serial & 1.2 min \\WHO & shapr iterative method & Parallel & 16 min
 \\WHO & shapr iterative method & Serial & 36 min\\
WHO & shapr sequential method & Parallel & 131 min \\
\hline

\end{tabular}\caption{We show the computing time for each dataset and method}
\label{tab:all_computing_times}
\end{table}

Our methods also suffer from the Curse of Dimensionality \citep{Bellman1961}: Even for high-performance computing, when $p$ increases, such as $p=40$, the machine eventually runs out of memory. Our machine runs out of memory for $p$ larger than about 21, but machines with more RAM can fit larger datasets. There is a constraint in R restricting the number of non-zero elements to be $2^{31}-1$. However, loading the spam-package \citep{spam} together with the R package spam64 \citep{spam64} enables the sparse matrix class spam to handle huge sparse matrices with more than $2^{31}-1$ non-zero elements. This is future work. There are other options available. One is sampling: instead of using a few hundred coalitions, as is often done, one can now use thousands. We have illustrated that model fitting will be fast, indicating that the new methods are a major improvement. Another option is to split the model estimation into equal ``chunks'', where for instance half of the models is fitted, then the remaining half is fitted and finally the results are combined before estimating the Shapley values.   

The Diagonal correction method should be contrasted to the so-called soft constraints \citep[p.~39]{rue2005gaussian}. Here the constraints are assumed to be observed with Gaussian noise. To obtain an equivalent result, we need to assume that the covariance for the noise term is the identity matrix with a small variance term included. Therefore the model is misspecified. The precision matrix of the field is then $\bm Q+\bm A^T\bm A\kappa$, but we need to assume that the observed error is $\bm 0$ to get the correct mean in our result. The soft constraint approximates the hard constraint under these assumptions, and the approximation becomes better and better the smaller the variance of the error term is. By using this result, we are assuming a slightly misspecified model, as the constraints are assumed to be observed with noise. In our result, we make no such assumptions. 

In comparison, our result tells us that by directly altering the precision matrix, we approximately perform kriging with hard constraints. By adding $\bm A^T\bm A\kappa$ to the precision matrix, we are forcing the precision matrix to have the directions in $\bm A$ as eigenvectors approximately. The variance in these directions becomes small. In general the directions in $\bm A$ do not need to be eigenvectors of $\bm Q$. However, the result justifies adding this correction term to the precision matrix.

The new methods could also be compared to the method of \cite{bolin2021efficient}. Here they also consider (proper) Gaussian models. One calculates a transformation matrix $\bm T$ that involves the constraints so that in the transformed space all constraints are enforced automatically. However, when there are $2^p$ different models, this involves calculating a large number of transformation matrices. This will slow the method down significantly compared to our methods.  

We have developed three new methods and compared them against each other and the two methods from shapr. The Exact method introduces no approximation. Based on the median absolute errors and run-times of the three methods, the methods perform similarly. Since there is no tuning parameter in the Exact transformation method, we suggest that the reader uses the Exact transformation method.
\bibliographystyle{chicago}
\bibliography{references.bib}

\end{document}